\documentclass[footinbib,a4paper,aps,prb,reprint,twocolumn,preprintnumbers,amsmath,amssymb,10pt, superscriptaddress]{revtex4-2}
\usepackage[english]{babel}
\usepackage{graphicx}
\usepackage[utf8]{inputenc}
\usepackage{color}
\usepackage{wrapfig}
\usepackage{ulem}
\usepackage{mathbbol}

\makeatletter
\renewcommand{\fnum@figure}{FIG. \thefigure}
\makeatother

\newcommand{\beq}{\begin{equation}}
\newcommand{\eeq}{\end{equation}}
\newcommand{\be}{\begin{equation}}
\newcommand{\ee}{\end{equation}}
\newcommand{\beqa}{\begin{eqnarray}}
\newcommand{\eeqa}{\end{eqnarray}}
\newcommand{\bea}{\begin{align}}
\newcommand{\eea}{\end{align}}

\newcommand{\mb}[1]{\mathbf{#1}}

\renewcommand{\rm}[1]{\textrm{#1}}
\newcommand{\CRO}{Ca$_2$RuO$_4$}
\newcommand{\CRObis}{Ca$_3$Ru$_2$O$_7$}
\newcommand{\ldw}{l_{\textrm{dw}}}
\newcommand{\Edw}{E_{\textrm{dw}}}
\newcommand{\Eel}{E_{\textrm{el}}}
\newcommand{\fel}{f_{\textrm{el}}}
\newcommand{\fb}{f_{\textrm{b}}}
\newcommand{\qmin}{q_{\textrm{min}}}

\newcommand{\Fel}{F_{\textrm{el}}}

\begin{document}

\title{Elastic stripes formation at electronic transitions}
\author{Giuliano Chiriacò}
\affiliation{The Abdus Salam International Center for Theoretical Physics, Strada Costiera 11, 34151 Trieste, Italy}
\affiliation{SISSA — International School of Advanced Studies, via Bonomea 265, 34136 Trieste, Italy}
\author{Andrew J. Millis}
\affiliation{Center for Computational Quantum Physics, The Flatiron Institute, New York, NY 10010, United States}
\affiliation{Department of Physics, Columbia University, New York, NY 10027, United States}

\date{\today}

\begin{abstract}
We investigate the possibility of a striped inhomogeneous phase occurring as an electronic system with an order parameter linearly coupled to the elastic degrees of freedom is tuned through the electronic phase transition. 
We find that in finite systems where boundary conditions create an elastic incompatibility,  a stripe pattern may emerge in the vicinity of the electronic transition. Stripes are stabilized when the gained elastic energy overcomes the typical electronic and domain wall energy costs. In thin film geometries, the stripes are found  to extend across the entire depth of the system. We also study the behavior of the phase fraction across the spinodal region of a first order phase transition. The orientation of stripes and the possibility of more complicated patterns are also discussed.
\end{abstract}

\maketitle

\normalem

\section{Introduction}

The metal-insulator transition in correlated systems is a topic of long standing interest in condensed matter physics \cite{MIT:review}. In many families of Mott materials exhibiting a metal-insulator transition, the electrons are coupled to the elastic degrees of freedom \cite{Mn:Hwang,Mn:Radaelli,Mn:Millis,Mercy:Ni,Pavarini:CRO,Ricco:CRO,SIO:1,SIO:2,VO2:str1,VO2:str2,VO2:str3,McLeo:Nat, Ben:Fra,ElMIT:Dominguez20,ElMIT:Georgescu19,ElMIT:Georgescu21,ElMIT:Qi21,ElMIT:Zhang16,ElMIT:Argama19,ElMIT:Rogge18,ElMIT:Schueller20,ElMIT:Uhlenbruck99,ElMIT:Zhang16,ElMIT:Zhaoliang18,HET:Chen21,HET:Disa17,HET:Hu16,HET:Phillips17,HET:Szymanski19,ElMIT:Lechermann20,Mattoni:2016,Suyolcu:2021,KIM:2017,McLeod:2016,PseudoStrain:2010}, and the electronic order at the transition is accompanied by a change in the lattice structure and by the appearance of elastic strain fields. Because of its important role in many materials, the coupling of elasticity to  both equilibrium and nonequilibrium metal insulator transitions has become a topic of active current interest. 

Recent experiments \cite{Gao:ref1,Gao:ref2,Gao:ref3,Kartha:martensite,CRO:237,Ahn:PRB,Ahn04,Mengkun:CRO,Stripes:Ronchi22} have revealed that in some cases the transition from uniform metallic to uniform insulating phase proceeds via an intermediate `stripe' phase characterized by an alternating mesoscale pattern of metallic and insulating domains. This behavior is believed \cite{Ahn04,VO2:str3,Gao:ref2,Mengkun:CRO,Gao:2002,Verri19} to arise from the interplay of the energy associated with long ranged strain fields (which  depends strongly on sample geometry) and the electronic energy that drives the phase transition. However, with a few exceptions that will be discussed below, most of the existing theoretical work focuses on spatially uniform situations, and the phenomenon of stripe formation is not understood in detail.

In an important study, Ahn and collaborators \cite{Ahn04} studied orbital ordering transitions in the ``colossal'' magnetoresistance manganites and showed that the coupling of orbital ordering to strain fields in the context of global geometrical constraints associated with clamped elastic boundary conditions could give rise to ``tweed'' domain patterns reminiscent of those observed at the martensite-austenite transition. These ideas were further developed by Guzman-Verri and collaborators \cite{Verri19}.

Gao and collaborators studied theoretically \cite{Gao:2002} and experimentally \cite{Gao:ref1,Gao:ref2,Gao:ref3} a system in which a monolayer adsorbed on the surface of a thick substrate material undergoes a phase transition that produces a strain field extending into the bulk of the substrate material. The strain field in the substrate costs elastic energy, which may be reduced by the formation of striped domains, which in turn increases the electronic energy cost. It was found that balancing these contributions leads to various non-uniform configurations, including parallel stripes and more complex bidimensional geometries. Similar ideas were applied to show that the metal-insulator interface induced by applying an electrical current to \CRO would in general be reconstructed into a stripe pattern \cite{Mengkun:CRO}.

In this paper we build on these ideas to provide a general theory of an electronic transition coupled to strain fields. While we are motivated by recent results on metal-insulator transitions, the precise nature of the transition is not relevant and the theory is generally applicable to any electronic transition coupled to elastic degrees of freedom. We write a general theory but focus our specific results on the typical case of samples in which the top surface presents a free boundary condition, while the bottom surface resting on the substrate is constrained to have a zero lattice displacement; this is the relevant case for a large class of experiments. We consider both modulations in the plane of the film (``stripes'') and variations over the depth of the film, generalizing recently studied models, which consider either a monolayer of potentially ordered phase \cite{Gao:2002}, or a very thin stripe layer with a thickness much smaller than that of the system \cite{Mengkun:CRO}.

The analysis is in general complicated, as the elastic theory describing actual low-symmetry materials involves many parameters, and the electron-lattice coupling parameter is a second rank tensor which can display anisotropies and shear stress components. The existence and nature of any inhomogeneous states will depend quantitatively on the precise elastic theory and electron-lattice coupling. However we can still make some general statements. Stripes may occur in systems where boundary conditions prevent an energetically favored spatially uniform elastic distortion. The emergence of stripes is primarily determined by the competition between the elastic energy gained from stripe formation and the energy cost of the electronic inhomogeneity caused by the stripes. These considerations suggest that stripes should be favored for temperatures close to the electronic phase transition temperature, where the energy cost of modulating the order parameter is small. This argument applies for both first and second order phase transitions; for first order transitions, stripes may be expected to  appear in a temperature range roughly spanning the interval between the upper and lower spinodals of the electronic transition. We also find that stripes extend over the full depth of the film and orient themselves in the direction that maximizes the elastic energy gain. 

Since the elastic interactions are long-range and have no intrinsic length scales, the periodicity of the stripes can only depend on the length scales set by the system: in our case these are the thickness of the sample and the bending lengthscale (i.e. the domain wall thickness) of the electronic order parameter. The orientation of the stripes is controlled by the anisotropies of the electron-lattice coupling tensor and the underlying elastic theory. There may be no preferred orientation (as we shall find for isotropic elasticity and planar-isotropic electron-lattice coupling) or two optimal orientations, such as in the presence of in-plane shear coupling. When two stripe orientations are possible, the system may form domains where one or the other orientation appear. Bistripe patterns (checkerboard-like domains with crossing non-parallel stripes) are possible in principle, but we find that they have higher energy because, while the elastic energy is the same as for unistripe patterns, they are more costly in terms of electronic energy.

The rest of this paper is organized as follows. In Section \ref{Model} we present the model and find the free energy by solving the elastic equations as function of the electronic order parameter and using the solution to  integrate out the lattice degrees of freedom.  In Section \ref{PT2} we calculate the total free energy for a system undergoing a second order phase transition, determine the stability of the stripe phase and show that bistripe patterns are energetically more costly. In Section \ref{PT1} we study the case of a first order phase transition, find the criteria for the appearance of stripes and calculate the behavior of the metallic and insulating phase fractions as function of the temperature within the spinodal region. In Section \ref{MetaStab} we consider the case of stripes extending only halfway through the sample and find that such phase is metastable. In Section \ref{Orientation} we study the preferred orientation of stripes and the effect of anisotropies on their stability. Finally in Section \ref{conclusions} we present our conclusions.

\section{General theory}\label{Model}

\subsection{Energy and formalism}
We consider a system with an electronic order parameter $\psi(\mb x)$, which in general depends on the position $\mb x$ and is linearly coupled to strain fields $\epsilon_{ij}$ so the total free energy is the sum of a purely electronic term $f_{e}$, an electron-lattice coupling $f_{\textrm{e-l}}$ and a lattice elastic energy $f_{\textrm{l}}$. We assume the elastic theory is harmonic. Thus the total free energy $F$ is given by
\begin{gather}\label{FTot}
F=\int d^3\mb x (f_{\textrm{e}}+f_{\textrm{l}}+f_{\textrm{e-l}});\\                                                 
\label{feff}  f_{\textrm{e-l}}\equiv-\sigma_{ij}\epsilon_{ij}\psi(\mb x);\qquad f_{\textrm{l}}\equiv\frac{1}{2}\epsilon_{ij}K^{ijlm}\epsilon_{lm},
\end{gather}
where $K^{ijlm}$ is the elastic tensor, $\epsilon_{ij}\equiv\frac12(\partial_iu_j+\partial_ju_i)$ is the strain tensor which depends on derivatives of the elastic displacement vector $u_i$, and $\sigma_{ij}$ is the stress mismatch tensor which quantifies the force exerted by the electronic order on the lattice and acts as a coupling parameter.

In the limit of a free infinite sample, we may minimize freely over the strain fields. Substituting the resulting $\epsilon_{ij}=\left(K^{ijlm}\right)^{-1}\sigma_{lm}\psi$ into Eq. \eqref{FTot} gives 
\begin{gather}\label{FTot1}
F=\int d^3\mb x \left(f_{\textrm{e}}+f_{\textrm{stab}}\right);\\
\label{feff1}   f_{\textrm{stab}}\equiv-\frac{\psi(\mb x)^2}{2}\sigma_{ij}\left(K^{ijlm}\right)^{-1}\sigma_{lm}.
\end{gather}
The term $f_{\textrm{stab}}$ is the energy gained by allowing the lattice to relax to the configuration optimized for a non-zero $\psi$ and plays an important role in the energetics of bulk transitions. Since the coefficient of $\psi(\mb x)^2$ is independent of $\mb x$, we find a spatially homogeneous phase.
\begin{figure}[t]
  \centering
  \includegraphics[width=0.9\columnwidth]{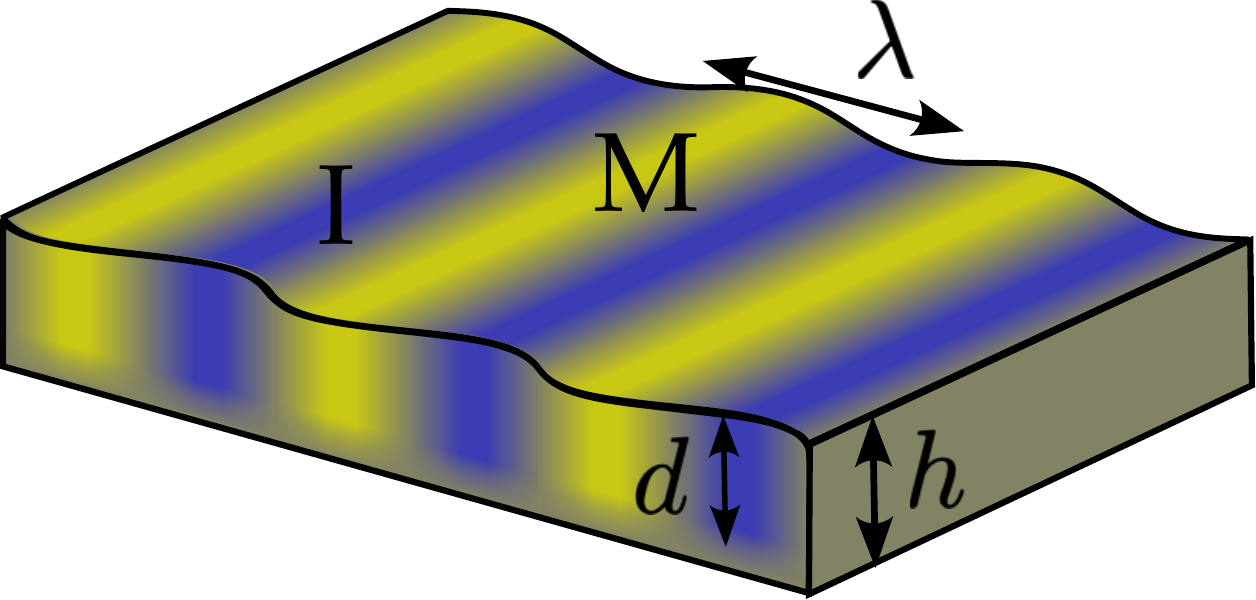}
  \caption{\scriptsize{Sketch of the system under consideration (with clamped boundary conditions at the bottom as an example). The metallic (M) and insulating (I) regions correspond to a different elastic behavior.}}\label{Fscheme}
\end{figure}

However, in many experimentally relevant situations boundary conditions constrain the ability of  the lattice  to relax to the minimum energy configuration. 
To analyze this situation more quantitatively we write  the Euler-Lagrange equations following from Eq. \eqref{FTot} for the strain fields $u$:
\begin{gather}\label{EUEQ}
\frac{\delta f_{\textrm{e}}}{\delta \psi(\mb x)}=\sigma_{ij}\epsilon_{ij};
\qquad
K^{ijkl}\partial_j\partial_ku_l=\sigma_{ij}\partial_j\psi.
\end{gather}

Conceptually, for each $\psi(\mb x)$ we solve the second Euler-Lagrange equation, imposing the appropriate boundary conditions on $u$, and from that compute the total free energy as a functional of $\psi(\mb x)$ only. In practice, this requires the knowledge of the elastic Green function for a finite system, whose calculation is often analytically cumbersome, so that it is simpler to start from a variational ansatz for $\psi$, explicitly calculate $u$ for such ansatz and then minimize $F$ with respect to the variational parameters describing $\psi$.

\subsection{Solution of the elastic equations}\label{ModelEqsSol}

In this subsection we present a formal solution of the elastic equation \eqref{EUEQ} assuming a fixed order parameter $\psi(\mathbf{x})$, for a system of size $L$ in the $x$ and $y$ directions and thickness $h$ in the $z$ direction,  assuming $L\gg h$ (see Fig. \ref{Fscheme}). The elastic deformation on the $x$-$y$ surfaces at $z=0$ and $z=h$  are constrained by boundary conditions. We consider the typical film geometry with one free surface (here $z=0)$ and one surface where the atomic displacements are clamped by coupling to a substrate ($z=h$). 

The boundary condition at the free surface can be found by integrating by parts Eq. \eqref{FTot} with respect to $u$ and minimizing the resulting boundary surface integral, leading to
\begin{equation}\label{BCK}
K^{ijkl}n_j\epsilon_{kl}=\sigma_{ij}n_j\psi.
\end{equation}
For a clamped surface, the boundary condition depends on the stiffness of the clamping substrate; for an infinitely stiff substrate, the displacement at the boundary has to vanish, i.e. $u_i=0$. The more complicated case of a substrate with a finite stiffness can be studied by considering the deformation of the substrate and imposing matching boundary conditions, but such analysis is not essential to capture the physics of the problem and is beyond the scope of our work.

Equation \eqref{EUEQ} can be solved in general but here we focus on the specific (and simpler) case of isotropic elasticity, which contains the essential physics. The elastic tensor is then given by $K^{ijlm}\equiv\Lambda\delta_{ij}\delta_{lm}+\mu\delta_{il}\delta_{jm}+\mu\delta_{im}\delta_{jl}$, with bulk modulus  $\Lambda\equiv E\nu/((1+\nu)(1-2\nu))$ and shear modulus $\mu\equiv E/(2(1+\nu))$, where $E$ is the Young modulus and $\nu$ the Poisson ratio \cite{LL7}. Equation \eqref{EUEQ} becomes
\begin{equation}\label{EqElIs}
(\Lambda+\mu)\partial_i\partial_ju_j+\mu\nabla^2u_i=\sigma_{ij}\partial_j\psi.
\end{equation}

We Fourier transform the order parameter in the $x$-$y$ plane
\begin{equation}\label{psiq}
\psi(\mb x)=\sum_{\mb q}e^{i\mb q\cdot\mb r}\psi_{\mb q}(z),
\end{equation}
where $\mb q\equiv(q_x,q_y)$ and $\mb r\equiv(x,y)$, and then solve Eq. \eqref{EqElIs} separately for each Fourier component $\psi_{\mb q}$.

The solution to Eq. \eqref{EqElIs} is the sum of a particular solution that accounts for the source term plus a homogeneous solution required to satisfy the boundary conditions:
\begin{gather}\label{usolT}
u_i(\mb x)=e^{i\mb q\cdot\mb r}\left(u^P_i(\mb q,z)+\sum_{\substack{a=0,1,2\\\pm}}A_a^{\pm}u^{\pm}_{a,i}(\mb q,z)\right);\\
\label{HomElEq} [(\Lambda+\mu)\partial_i\partial_j+\mu\nabla^2](e^{i\mb q\cdot\mb r}u^{\pm}_{a,i})=0,
\end{gather}
where $A_a^{\pm}$ are constant coefficients determined by the six boundary conditions (three at each interface). The solutions of the homogeneous equations are
\begin{gather}
\label{uHom0}u_0^{\pm}=e^{\mp qz}\left[\begin{pmatrix}1\\0\\0\end{pmatrix}\mp\frac{q_xz}{3-4\nu}\begin{pmatrix}q_x/q\\q_y/q\\\pm i\end{pmatrix}\right];\\
\label{uHom1}u_1^{\pm}=e^{\mp qz}\begin{pmatrix}q_x/q\\q_y/q\\\pm i\end{pmatrix};\qquad
u_2^{\pm}=e^{\mp qz}\begin{pmatrix}q_y/q\\-q_x/q\\0\end{pmatrix},
\end{gather}
with $q\equiv|\mb q|$.

If the top surface of the system $z=0$ is free of any external stress, and the bottom surface $z=h$ is clamped, the boundary conditions are
\begin{gather}
\label{BC01}\frac{E}{2(1+\nu)}(\partial_xu_z+\partial_zu_x)_{|z=0}=\sigma_{xz}\psi_{\mb q}(0);\\
\label{BC02}\frac{E}{2(1+\nu)}(\partial_yu_z+\partial_zu_y)_{|z=0}=\sigma_{yz}\psi_{\mb q}(0);\\
\label{BC03}E\frac{(1-\nu)\partial_zu_z+\nu(\partial_xu_x+\partial_yu_y)}{(1+\nu)(1-2\nu)}|_{z=0}=\sigma_{zz}\psi_{\mb q}(0);\\
\label{BCh}u_x(z=h)=0;\quad u_y(z=h)=0;\quad u_z(z=h)=0.
\end{gather}

Note that the boundary conditions at $z=0$ \eqref{BC01}-\eqref{BC03} still display the presence of a nonzero stress, which is ``internal'' since it is related to coupling with the electrons.

Combining Eqs. \eqref{uHom0}-\eqref{BCh} we can solve for $A_a^{\pm}$ and then substitute back into Eq. \eqref{FTot} to find the elastic energy contribution (the exact procedure is outlined in the Appendix \ref{App:A}).

\subsection{Thickness and $z$ profile of the stripes}\label{Thickness}

The depth profile ($z$-dependence) of $\psi_{\mb q}(z)$ determines the particular solution $u^P$, thus affecting the homogeneous $u^H$ solution and the elastic energy.

In this subsection we give a heuristic argument why the energy is lower when stripes extend over the entire thickness of the film. The basic idea is that the energy gain may be written as $-\int \epsilon(\mb x)\sigma\psi(\mb x)$, as it comes only from the regions where the order parameter is non-zero; one expects that this region should be made as large as possible.

Indeed, if we assume the order parameter extends over a typical distance $1/\kappa$, the particular solution will also extend over $1/\kappa$. Furthermore, for a stripe of wavelength $q$, the homogeneous solution and its associated strain extend over a distance $1/q$ away from the boundaries; the strain energy cost is therefore $\sim 1/q$ while we may write the energy gain as $\sim 1/\kappa f(q/\kappa)$. A quick minimization yields a $q\sim \kappa$ and thus a total energy gain $\sim 1/\kappa$, implying that $\kappa^{-1}$ should be made as large as possible.

This finding is rather different from the theoretical model of current induced stripes in \CRO \cite{Mengkun:CRO}, where the stripes have a smaller optimal wavelength, $q_{\textrm{min}}\sim1/d$, and are confined to a thin surface layer $d\ll h$. This is due to the different mechanism responsible for the transition, which is driven by an electric current flowing close to the surface of the system. This causes an inhomogeneous temperature variation \cite{CM:pol} and thus a spatially inhomogeneous free energy landscape, which favors stripes with a smaller thickness and wavelength.

In addition to this heuristic argument, we also note that stripes extending only halfway through the film incur in an additional energy cost by creating in-plane domain walls between metallic and insulating domains that are not present when $d=h$. Nonetheless, it is possible for a phase with stripes extending down to a depth $d<h$ to exist in a metastable form, although at much higher energy. This case is addressed in Section \ref{MetaStab}, while in all other sections we consider stripes extending uniformly over the entire thickness of the film.

Therefore, except for Section \ref{MetaStab}, we assume an order parameter independent of $z$. We remark that while this is not a locally exact solution of Eq. \eqref{EUEQ}, since $\epsilon_{ij}$ will in general have a $z$ dependence, it is a good variational approximation of the true solution, in the sense that our procedure produces an approximate solution with a free energy very close to the true minimum.

\subsection{Elastic free energy}\label{ModelElastic}

For stripes extending over the entire depth of the film the order parameter is independent of $z$ and we can write the elastic contribution to the free energy in the form
\begin{gather}
\notag F_{\textrm{el}}\equiv\int d\mb x(f_{\textrm l}+f_{\textrm{e-l}})=-\frac12\int d\mb x\sigma_{ij}\epsilon_{ij}\psi;\\
\label{FEl}F_{\textrm{el}}=\Eel\int d\mb x\sum_{\mb q}f_{\textrm{el}}(\mb q)|\psi_{\mb q}|^2,
\end{gather}
where $\Eel\equiv\sigma_{xx}^2/2E$ and $\fel$ is a dimensionless function of $qh$, $\theta\equiv\tan^{-1}(q_y/q_x)$ and $\sigma_{ij}/\sigma_{xx}$.

The analytic expression for $\fel$ is rather cumbersome and is derived and written explicitly in Appendix \ref{App:A}. This makes the minimization of $\fel$ with respect to $\theta$ and $q$ hard and it is not straightforward to determine a priori the favored orientation and periodicity of the stripes, but we can make some general considerations.

Since elasticity is essentially a long-range interaction, there is no lengthscale in the elastic equations except for the thickness $h$  \footnote{Here we are assuming that $L\gg h$, or that the periodicity of the stripes occurs on a scale much smaller than the size of the sample in the $x$ direction, which appears to be the case in all the relevant experiments.}. Thus if $\fel$ displays a minimum at finite $q$ its position only depends on $h$: for typical values of $\nu$ the minimum is located around $qh\sim2$.
\begin{figure}[t!]
  \centering
  \includegraphics[width=\columnwidth]{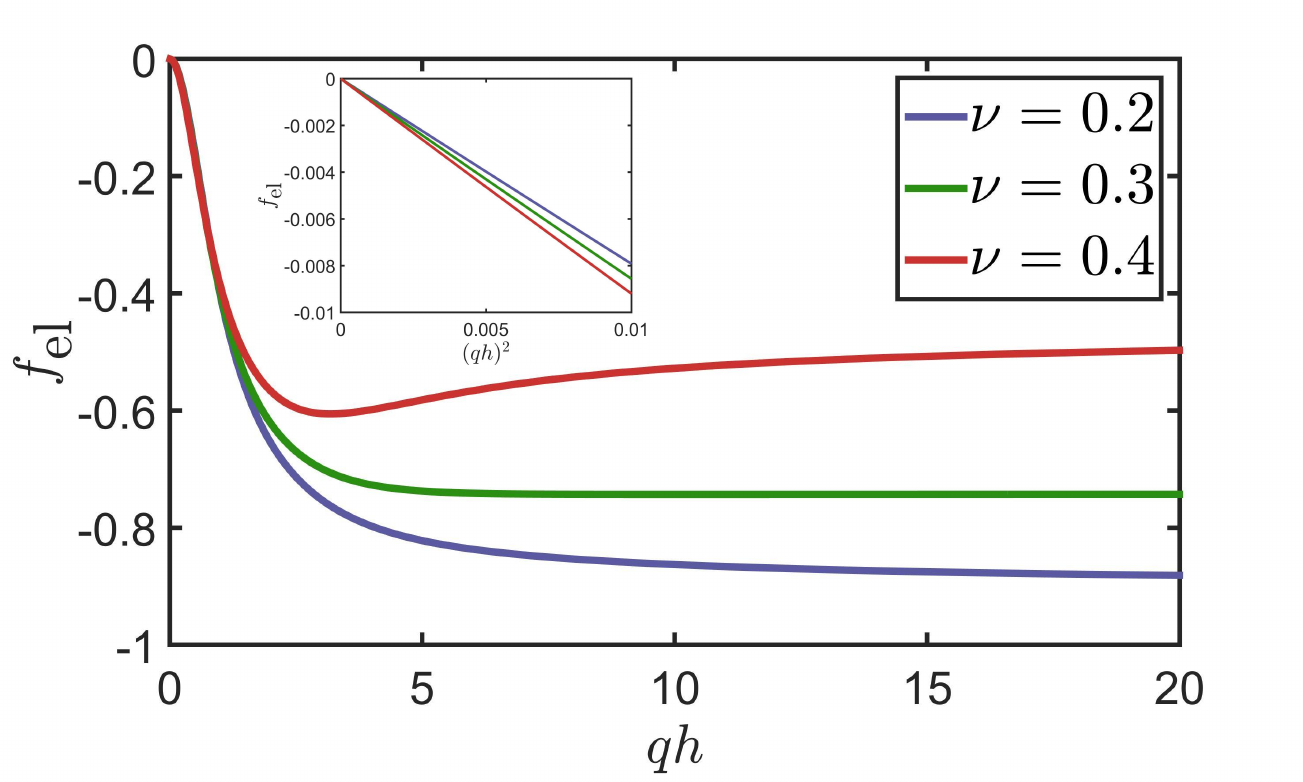}
  \caption{\scriptsize{Plot of $f_{\textrm{el}}$ for three values of $\nu$, $\sigma_{xx}=\sigma_{yy}$ and the remaining $\sigma_{ij}=0$; the inset shows $f_{\textrm{el}}$ as function of $(qh)^2$ at small values of $q$, displaying the quadratic behavior of the elastic energy near $q=0$.}}\label{F1}
\end{figure}

Therefore we can distinguish three qualitatively different cases. \emph{(i)} The minimum of $\fel$ is at $q=0$: stripes are not energetically favored and can never appear. \emph{(ii)} The minimum of $\fel$ is located at finite $qh$. \emph{(iii)} The minimum of $\fel$ is located at $q=\infty$. We remark that in case \emph{(ii)} and \emph{(iii)}, stripes are not guaranteed to appear as they still have to compete with the energy cost of forming domain walls. However, if they appear, their periodicity in case \emph{(iii)} is determined by a microscopic lengthscale such as the typical domain wall thickness, while in case \emph{(ii)} it is determined by the thickness $h$. As we will see in Section \ref{PT2}, the appearance of stripes is determined by the coefficient of the quadratic term of $\fel(q)$ at small $q$.

The  orientation of stripes relative to the atomic axes of the material  is also an important question but is not essential  to the study of the physical mechanism related to the formation of stripes, and the criteria for their appearance. For convenience, we postpone a detailed discussion of the orientation of stripes to Section \ref{Orientation}, where we shall discuss the main symmetry classes of $\sigma_{ij}$ and address in detail the effect of in-plane shear coupling $\sigma_{xy}$ as well as out of plane shear $\sigma_{xz}\neq0$. In the next sections we consider the simple planar isotropic case, $\sigma_{xx}=\sigma_{yy}=\sigma$ and all the other components $\sigma_{ij}=0$. Such isotropy in the $x$-$y$ plane means there is no preferred orientation of the stripes, and without loss of generality we can consider the case $\mb q=(q,0)$.

Since the order parameter is homogeneous in the $z$ direction, i.e. $\psi_{q}(z)\sim\text{const}$, the particular solution for each Fourier component is found at once from Eq. \eqref{EqElIs}:
\begin{equation}\label{uPdh}
\vec{u}^P=\hat x\frac{\psi_q}{iq}\frac{\sigma_{xx}2(1+\nu)}E\frac{1-2\nu}{2(1-\nu)}.
\end{equation}

After imposing the boundary conditions for each Fourier component and integrating out $u_i$, the dimensionless elastic free energy is found, see Eqs. \eqref{Eq:FEL}-\eqref{Eq:fxx}. To study the appearance of stripes we do not need the exact expression of $f_{\textrm{el}}$, but only its asymptotic behaviors, which we report here:
\begin{gather}
\label{FhasymInf}\frac{f_{\textrm{el}}(q\rightarrow\infty)}{1+\nu}=
-\frac{1-2\nu}{1-\nu}+\frac{2(1-4\nu+\nu^2+4\nu^3)}{(1-\nu)(3-4\nu)qh};\\
\label{Fhasym0}\frac{f_{\textrm{el}}(q\sim0)}{1+\nu}=-c_{q^2}(qh)^2;\qquad c_{q^2}\equiv\frac23.
\end{gather}

The free energy vanishes at $q=0$ and behaves quadratically with a negative coefficient for small $q$. For $q\rightarrow\infty$ it goes to a finite value that depends on $\nu$, with an asymptotic behavior $\sim1/q$. Since $c_{q^2}>0$, in this regime the minimum of $\fel$ is never at $q=0$; depending on $\nu$ the minimum of $\fel$ occurs at infinite or finite $q\neq0$.

\section{Second order phase transition}\label{PT2}

In this section we study a system exhibiting an electronic phase transition of second order, at a critical temperature $T_c$. We describe the electronic free energy $f_{\textrm e}$ with the usual quartic order expression
\begin{equation}\label{fe2PT}
f_{\textrm e}=f_0\left(\frac{\psi^4}4+\alpha(T)\frac{\psi^2}2+\frac{\xi^2}2|\nabla\psi|^2\right),
\end{equation}
where $\alpha(T)=(T-T_c)/T_c$ near the transition and $\xi$ is a lengthscale associated to domain walls (and determined by the energy cost due to an inhomogeneous order parameter).

The contribution of the elastic energy \eqref{FEl} renormalizes the quadratic term in $\psi$, which we write in momentum space
\begin{gather}
\label{Ftot2PT}F=f_0\int d^3\mb x\frac{\psi^4}4+\frac12hL^2f_0\sum_q\alpha'(q,T)|\psi_q|^2;\\
\label{alfa1}\alpha'(q,T)\equiv\alpha(T)+\xi^2q^2+\frac{2\Eel}{f_0}f_{\textrm{el}}(q).
\end{gather}

When the system is close to the transition, we do not need to make a variational ansatz for $\psi$. In fact, we can study the instability of the system to an ordered ($\psi\neq0$) phase simply by looking at the sign of the coefficient $\alpha'$. If $\alpha'(q,T)<0$ an ordered phase is favored compared to the disordered phase $\psi=0$. In particular the minimum of $\alpha'$ with respect to $q$ determines which mode develops the instability first, i.e. at higher temperatures:
\begin{equation}\label{Tc}
\frac{T_c'}{T_c}=1-\left(\xi^2q_{\rm{min}}^2+2\frac{\Eel}{f_0}\fel(q_{\rm{min}})\right).
\end{equation}

\begin{figure}[t!]
  \centering
  \includegraphics[width=\columnwidth]{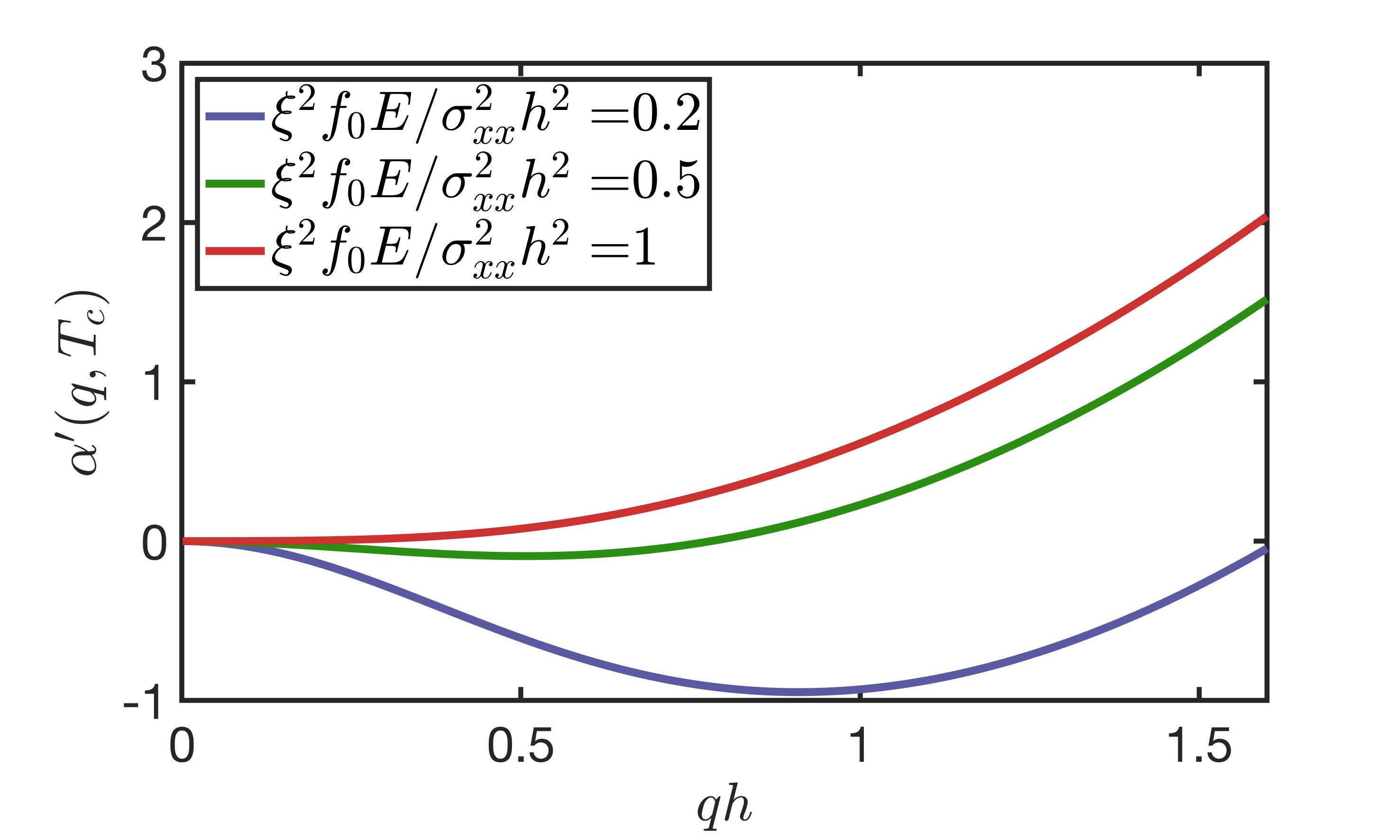}
  \caption{\scriptsize{Plot of $\alpha'(q)$ as function of $qh$ for three different values of $f_0E/\sigma_{xx}^2\xi^2/h^2$ and for $\nu=0.4$.}}\label{F3}
\end{figure}

Thus the system exhibits an inhomogeneous ordered phase with a finite wavevector $q_{\textrm{min}}\neq0$ if $\alpha'(q_{\textrm{min}})<\alpha'(0)$. Since the function $f_{\textrm{el}}(q)$ is limited, the $\xi^2q^2$ positive term dominates at large $q$; moreover $f_{\textrm{el}}(q)$ behaves like $\sim q^2$ near $q=0$, and thus the magnitude of its second derivative is crucial in determining where the minimum of $\alpha'(q)$ is located. Indeed when
\begin{equation}\label{2PTthr}
\xi^2+\frac{\Eel}{f_0}f''_{\textrm{el}}(q=0)<0\,\Rightarrow\, f_0\frac{\xi^2}{h^2}<2\Eel(1+\nu)c_{q^2},
\end{equation}
then the concavity of $\alpha'$ at $q=0$ is negative and there is an instability at $q_{\textrm{min}}\neq0$, see Fig. \ref{F3}. This makes sense, since the inhomogeneous striped order arises when the coupling between electrons and lattice degrees of freedom is strong enough to overcome the typical energy cost of creating a domain wall.

Using Eq. \eqref{Fhasym0} we find the quantitative criterion for a stripe pattern to form:
\begin{equation}
\frac{\sigma_{xx}^2}E>\frac{3}{2(1+\nu)} f_0\frac{\xi^2}{h^2}.
\end{equation}
Notice the role played by the thickness of the sample $h$: thin films require a stronger coupling $\sim\sigma_{xx}$ to develop an inhomogeneous phase. This is reasonable, because in thinner systems the energy cost of elastic displacement is higher since they have ``less depth'' to accommodate such displacement, which has to vanish at $z=h$ because of the clamped boundary conditions we employ.

We remark that the appearance of the striped phase changes the transition temperature from the ``uncoupled'' critical value $T_c$ to a ``true'' (i.e. observed) critical temperature $T_c'$, which is larger than $T_c$ given the stabilizing effect of the elastic degrees of freedom on the ordered phase.

We make a final observation on the possibility of having a checkerboard pattern, i.e. an order parameter of the type $\psi_{cb}\sim(\psi_xe^{iq_xx}+\psi_ye^{iq_yy})$, as opposed to simple stripes $\psi_{s}\sim\psi_0e^{iqx}$. For simplicity we make the assumption that these two types of profiles are described by plane waves; the exact form of $\psi(\mb x)$ may be more complicated but at a zeroth order approximation the qualitative physics is captured by this assumption. The optimal value of $q$ is determined by the minimum of Eq. \eqref{alfa1}: we consider the isotropic case ($\sigma_{xx}=\sigma_{yy}$), so that for the checkerboard pattern the minimization operates separately on $q_x$ and $q_y$, and we find $q_x=q_y=q$. Therefore, the magnitude of the periodicity is the same for both stripes profiles. 

The optimal values of $\psi_x$, $\psi_y$ and $\psi_0$ are determined by the competition between the quartic and quadratic terms in Eq. \eqref{Ftot2PT}. Substituting for the stripes pattern we find $F_s=f_0hL^2(\psi_0^4+\alpha'_{\textrm{min}}\psi_0^2/2)$, where $\alpha'_{\textrm{min}}<0$ is the value of $\alpha'$ at its minimum. The total energy is then minimized by $\psi_0=\sqrt{-\alpha'_{\textrm{min}}}$ and is $F_s=-f_0hL^2(\alpha'_{\textrm{min}})^2/4$.
\begin{figure}[t!]
  \centering
  \includegraphics[width=0.96\columnwidth]{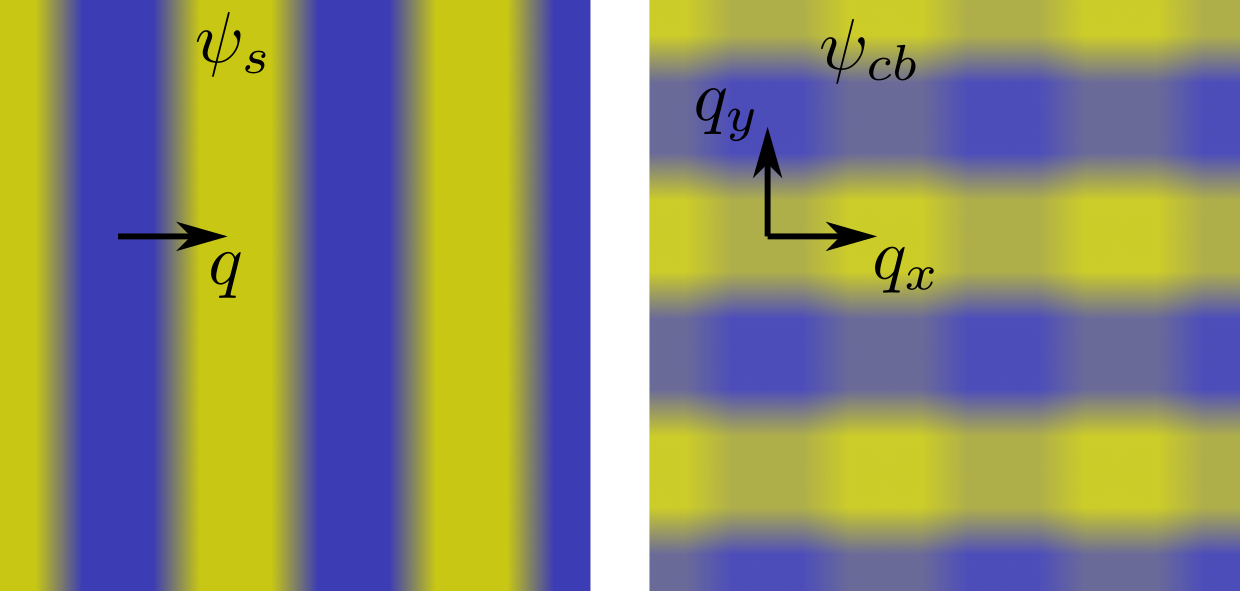}
  \caption{\scriptsize{Sketch of the stripes (left) and checkerboard profile in the $x-y$ plane; the yellow (blue) color indicates a metallic (insulating) phase.}}\label{F:checkerboard}
\end{figure}

Performing the same calculation for the checkerboard profile and using that $|\psi_{cb}|^2=\psi_x^2+\psi_y^2+2\psi_x\psi_y\cos(qx-qy)$, we find
\begin{gather}
\notag F_{cb}=f_0\int d^3\mb x\frac{|\psi_{cb}^4|}4+\frac12hL^2f_0\alpha'_{\textrm{min}}(\psi_x^2+\psi_y^2);\\
\label{Eq:Fcb}\frac{F_{cb}}{hL^2f_0}=\frac{(\psi_x^2+\psi_y^2)^2+2\psi_x^2\psi_y^2}4+\frac{\alpha'_{\textrm{min}}}2(\psi_x^2+\psi_y^2);
\end{gather}

Minimizing with respect to $\psi_x$ and $\psi_y$ we find $\psi_x^2=\psi_y^2=-\alpha'_{\textrm{min}}/3$, resulting in an energy $F_{cb}=-f_0hL^2(\alpha'_{\textrm{min}})^2/6>F_s$. Therefore a checkerboard pattern has a higher energy compared to a stripe pattern due to the mixed term $2\psi_x^2\psi_y^2$, which essentially correspond to the regions where $|\psi_{cb}|^2$ oscillates like $\psi_x^2\psi_y^2\cos(qx-qy)$, and where the order parameter substantially deviates from what would be its optimal value.

This argument applies also to other choices of electronic free energy: in general the system wants to have regions of pure metal or regions of pure insulator, while a checkerboard pattern also presents ``mixed'' regions where $\psi$ assumes energetically costly values in between the metallic and insulating values of $\psi$, see Fig. \ref{F:checkerboard}. We note that compared to a system free to relax, where there are distortions along both $x$ and $y$, the presence of boundary conditions and the geometry of stripes reduce the stabilization effect, since it constrains the distortions to be along one direction only, thus reducing the elastic energy gained.

\section{First order phase transition}\label{PT1}

In this section we study the case of a first order phase transition.

\begin{figure}[t!]
    \centering
    \includegraphics[width=\columnwidth]{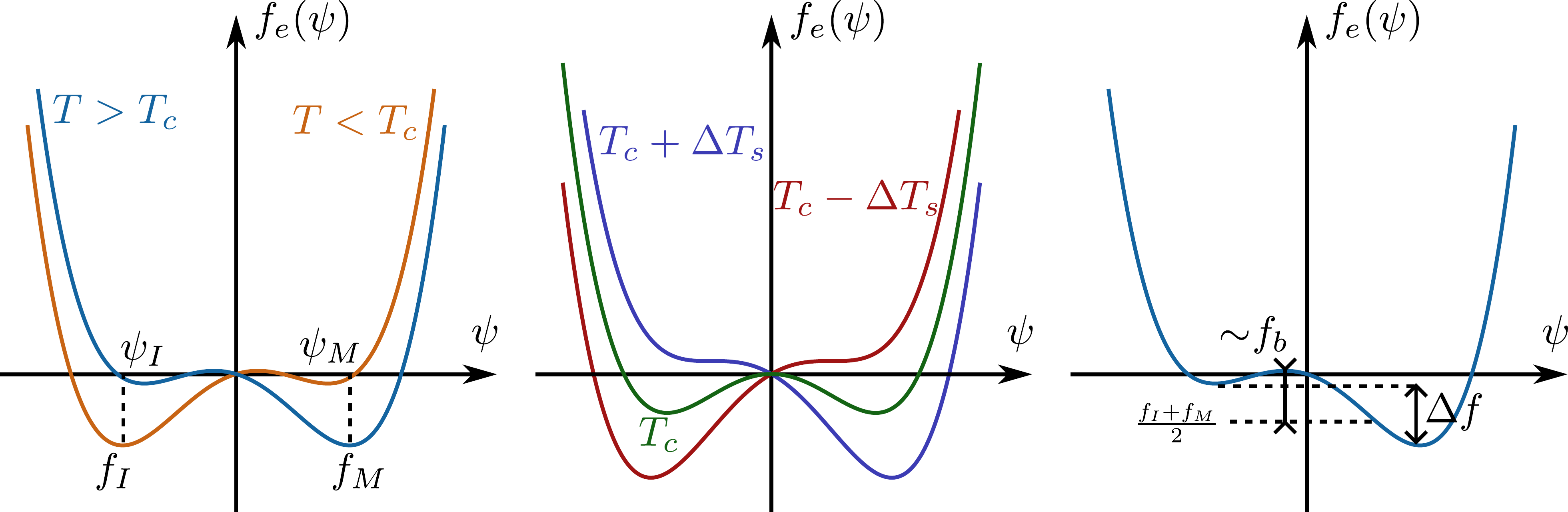}
    \caption{\scriptsize{Behavior of the electronic free energy $f_e$ as function of the order parameter $\psi$ for different values of the temperature. Left: behavior for $T_c<T<T_c+\Delta T_s$ (light blu) which favors a metallic phase with minimum value $f_M$ at $\psi_M$, and for $T_c-\Delta T_s<T<T_c$ (orange) which favors an insulating phase with minimum value $f_I$ at $\psi_I$. Middle: limiting cases at $T_c\pm\Delta T_s$ and $T_c$. Right: graphic estimate of the energy barrier $f_b$ and of the phase energy difference $\Delta f$.}}
    \label{F:fe}
\end{figure}

In general, systems undergoing a first order phase transition exhibit a coexistence of the two phases within the spinodal region of temperatures, in which both the metallic and insulating phase are locally stable. The free energy $f_{\textrm{e}}$ exhibits two minima at $\psi_M$ (metallic phase) and $\psi_I$ (insulating phase); outside of the spinodal region only one phase exists (either metal or insulator depending on the temperature). The two minima have energies $f_M\equiv f_{\textrm e}(\psi_M)$ and $f_I\equiv f_{\textrm e}(\psi_I)$, and in between them there is a local maximum of $f_{\textrm{e}}$, which creates an energy barrier $f_{\textrm b}$ that has to be overcome when switching phases, see Fig. \ref{F:fe}.

The phase energy difference $\Delta f\equiv f_M-f_I$ is crucial in determining the favored phase \footnote{Note that $f_{\textrm b}$ and $\Delta f$ are different, since the former is a sort of activation energy required to switch between $\psi_M$ and $\psi_I$, while the second is only the difference in energy between the two phases; in general $f_{\textrm b}>\Delta f$.}. At the critical temperature $T_c$ the two phases have the same energy, while the metal phase is favored above the critical temperature, so that $\Delta f$ vanishes at $T_c$ and has the opposite sign of $T-T_c$. The exact dependence of $\Delta f$ on $T$ is model dependent; for simplicity we assume that $\Delta f$ is approximately linear in $T$. Such a dependence may arise for example from this simple model of Landau free energy
\begin{gather}\label{fechoice}
f_{\textrm e}=f_0\left(\frac{\psi^4}{4}-\frac{\psi^2}2+g(T)\psi+\xi^2|\nabla\psi|^2\right);\\
\notag g(T)=\frac{2}{3\sqrt3}\frac{T_c-T}{\Delta T_s}.
\end{gather}
We have included the gradient contribution $\sim|\nabla\psi|^2$ to the free energy (with $\xi$ again a typical domain wall lengthscale) and defined the spinodal region by $T_c-\Delta T_s<T<T_c+\Delta T_s$; $f_0$ defines the scale of the electronic free energy.

\subsection{Variational ansatz for the order parameter}\label{PT1OP}

We now formulate a variational ansatz for the order parameter.

Deviations of $\psi$ from the optimal values $\psi_M$ and $\psi_I$ are typically very costly in energy, since they contribute in two ways: through the gradient term $|\nabla\psi|^2$ and through the energy barrier $f_{\textrm b}$ associated to the unfavorable values of $\psi$ between $\psi_I$ and $\psi_M$. Therefore, we expect the order parameter to exhibit very thin domain walls and have the shape of a square wave that switches abruptly between $\psi_M$ and $\psi_I$, with the width $\ldw$ of the switching region being very small compared to the wavelength $\lambda$ of the stripes. We consider stripes that are periodic in the $x$ direction and extend homogeneously in the $z$ direction.

We indicate with $q$ the wavevector describing the periodicity of the stripes, and write our ansatz for the order parameter as
\begin{gather}\label{OPpsi}
\psi(\mb x)=\psi_I+\Delta\psi\sum_n\phi(x-n\lambda);\\
\notag\phi(s)\equiv\tanh(s/\ldw)-\tanh((s-\eta\lambda)/\ldw),
\end{gather}
where $\lambda\equiv2\pi/q$, $\Delta\psi\equiv(\psi_M-\psi_I)/2$ and $\eta$ is the metallic phase fraction, i.e. the ratio of surface area occupied by the metal phase to the total surface area of the system \footnote{Note that when $d=h$, $\eta$ is also the mtallic volume fraction. However the definition in terms of surface fraction is easier to generalize to the $d<h$ case.}. We use $\phi$ as a function that regularizes the jumps between $\psi_M$ and $\psi_I$; the quantitative choice of hyperbolic tangent functions does not affect the end result. 

The variational parameters with respect to which we minimize the free energy are $q$, $\eta$ and $\ldw$.

\subsection{Electronic energy}\label{PT1En}

We now want to calculate the electronic contribution to the free energy arising from the order parameter ansatz. It can be split into three terms: one coming from the metallic regions where the order parameter is constant $\psi=\psi_M$, one coming from the insulating regions $\psi=\psi_I$ and one arising from the domain wall regions in which $\psi$ switches between $\psi_I$ and $\psi_M$.

The first term contributes with an energy $f_M$ over a volume fraction $\eta$ (which is the fraction of volume of the system occupied by the metallic phase), the second term contributes $f_I$ over a volume fraction $(1-\eta)$.

The domain walls contribute with a term proportional to the width $\ldw$ of the switching region, to the depth $h$ and to the energy barrier $\fb$ between the minima \footnote{In a more rigorous way we define $\fb$ as $\frac1{\ldw}\int_{-\ldw}^{\ldw}f(\psi(x))dx-\frac12(f_I+f_M)$.}; they also contribute through the gradient term
\begin{equation}\label{Fgr}
\int d\mb x f_0\xi^2|\nabla\psi|^2=L^2\frac{f_0\xi^2}{\ldw}\Delta\psi^2qh,
\end{equation}
since the number of domain walls is $\sim L/\lambda\sim qL$ and they extend over a length $\ldw$ along $x$, $L$ along $y$ and $h$ along $z$. We can also absorb any numerical coefficient arising from the integration of $|\nabla\psi|^2$ into the definition of $\xi$ and thus the total domain wall free energy is
\begin{equation}\label{Fgr2}
\frac{F_{\textrm{dw}}}{hL^2}=q\ldw f_{\textrm b}+\frac{f_0\xi^2}{\ldw}\Delta\psi^2q.
\end{equation}

We observe that we can minimize the free energy given by Eq. \eqref{Fgr2} with respect to $\ldw$ at once, and find that the domain wall thickness is given by $\ldw=\Delta\psi\xi\sqrt{f_0/f_{\textrm b}}$. Since $f_0$ and $f_{\textrm b}$ have the same order of magnitude, we find unsurprisingly that $\ldw\sim\xi\ll h$.

The total electronic free energy density is then
\begin{gather}\label{Fetot}
\frac{F_{\textrm{e}}}{hL^2}=f_I\left(1-\eta\right)+f_M\eta+2q\Delta\psi\xi\sqrt{f_0f_{\textrm b}}.
\end{gather}

\subsection{Formation of stripes and temperature dependence}\label{PT1T}
\begin{figure}[t!]
    \centering
    \includegraphics[width=\columnwidth]{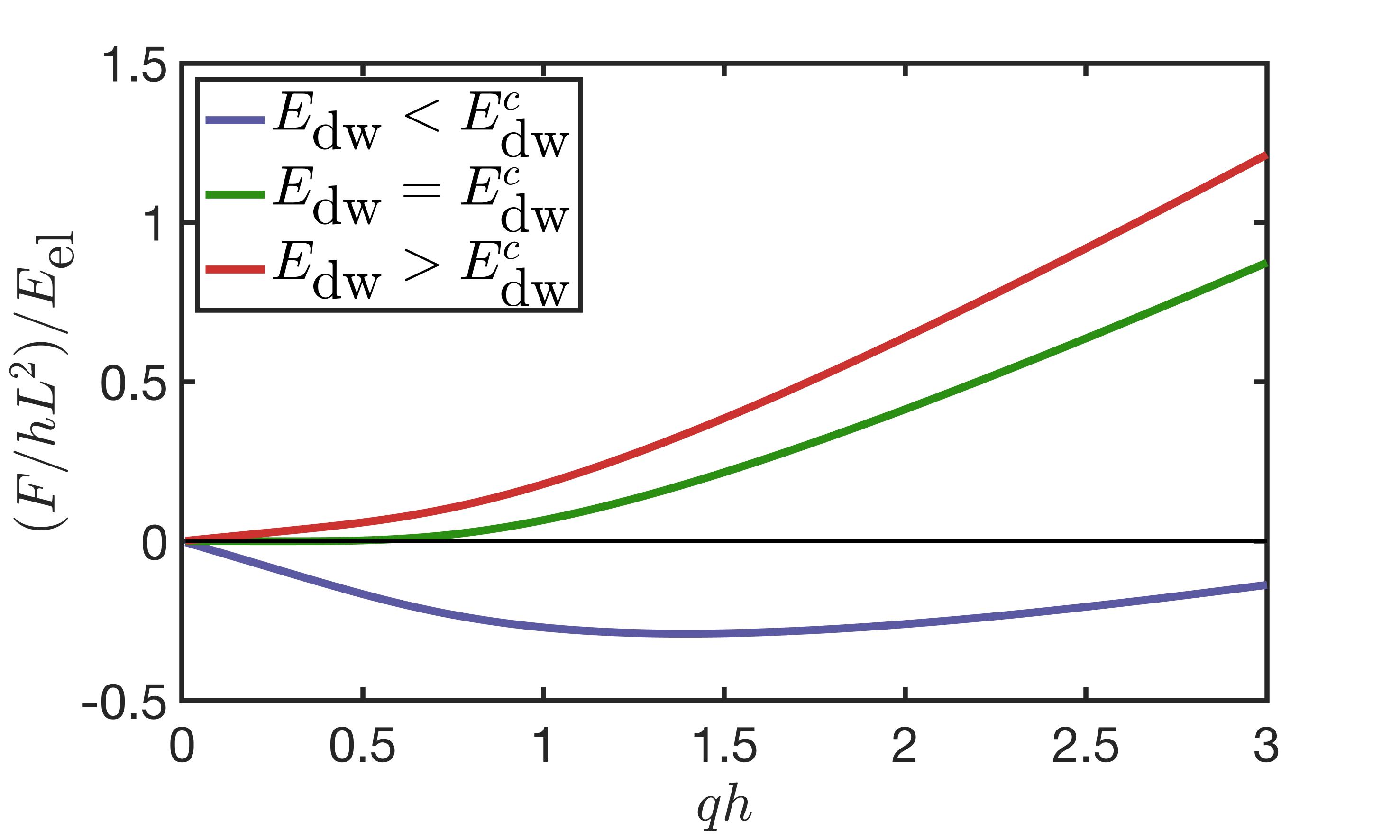}
    \caption{\scriptsize{Plot of the total free energy density normalized to $\Eel$ as function of $qh$ for $\Delta f=0$, $\nu=0.4$ for $\Edw=\Edw^c=0.49\Eel$, and $\Edw<\Edw^c$ and $\Edw>\Edw^c$.}}
    \label{F4}
\end{figure}

Combining Eqs. \eqref{FEl}, \eqref{OPpsi} and \eqref{Fetot}, the total energy can be written as
\begin{equation}\label{FT}
\frac F{hL^2}=\Edw qh+\eta\Delta f+E_{\textrm{el}}\sum_mf_{\textrm{el}}(q_m)|\psi_{q_m}|^2,
\end{equation}
where we dropped the constant term $f_I$ and defined the domain wall energy scale $\Edw\equiv2\frac{\xi}{h}\Delta\psi\sqrt{f_0f_{\textrm b}}$. 

Since $\ldw q\ll1$ we have approximated the order parameter with a square wave and Fourier transformed it, writing $\Fel$ as a sum over integer multiples of $q$ (i.e. $q_m=mq$ with $m$ integer); the Fourier component is $|\psi_{q_m}|^2=2\Delta\psi^2\sin^2(\pi\eta m)/\pi^2m^2$.

In general, the sum over $q_m$ has to be performed numerically, but we can make some observations: $\Fel$ is exactly zero for $\eta=0$ and $\eta=1$ (no elastic energy is gained when the system is homogeneous), and is symmetric around $\eta=1/2$ where it exhibits a minimum (the elastic energy gain is maximized for equal phase fraction).

We can find the analytic behavior of $\Fel$ at small and large $q$. For $qh\gtrsim1$, $\fel$ rapidly converges to its asymptotic value, so that the sum can be well approximated by a constant: $\Fel\sim\fel(q\rightarrow\infty)\sum_{m}|\psi_{q_m}|^2\sim\fel(q\rightarrow\infty)$. For small $q$, the elastic energy is quadratic and all terms such that $q_mh\lesssim1$ have a comparable contribution $\sim q_m^2\sin^2(\pi\eta m)/m^2$; thus the main contribution to the sum comes from these wavevectors $\sum_{q_mh\lesssim1}q_m^2\sin^2(\pi\eta m)/m^2\sim\sum_{m\lesssim1/qh}q^2/2\sim q$:
\begin{equation}\label{FsumAp}
\Fel\sim\Eel\fel''(q=0)q/h
\end{equation}
i.e. the summed free energy is now linear in $q$ at small $q$, rather than quadratic; we approximated the $\sin^2$ factor with its average since we are summing over many values of $m$ (see also Appendix \ref{App:sumfel}).

The domain wall energy is also linear in $q$, so that the total energy is positive for $q\rightarrow\infty$. Therefore the minimum  of the total energy is at $q=0$ when the slope of the elastic energy near $q=0$ is smaller in magnitude than the slope of the domain wall energy: up to some numerical factor coming from the sum over $q_m$ we can write this condition from Eqs. \eqref{Fhasym0}, \eqref{FT}-\eqref{FsumAp} as $\Edw\geq\Edw^c\sim\Eel$. Therefore the system exhibits a stripe phase with periodicity given by the wavevector $\qmin\sim1/h$ when the elastic energy is large enough compared to the typical domain wall energy, i.e. when
\begin{equation}\label{EdwcC}
\Edw/E_{\textrm{el}}\lesssim1\,\,\leftrightarrow\,\,\sigma_{xx}^2\gtrsim E\sqrt{f_0f_b}\xi/h.
\end{equation}
This criterion is very similar to what we found in Section \ref{PT2} for a second order phase transition, except for the dependence on the energy barrier $\fb$ (typical of a first order transition but absent in a second order transition). The thickness of the system again plays an important role, with stripe ordered phases being suppressed in thin films (smaller $h$).

Also notice that the criterion \eqref{EdwcC} is independent of $\eta$ and the temperature $T$, since as $\Edw\rightarrow\Edw^c$, $\qmin\rightarrow0$ and the approximations leading to Eq. \eqref{EdwcC} are exact. Indeed, when $\Edw\geq\Edw^c$ the energy contribution from elasticity and from domain walls goes exactly to zero (since there are no stripes) and the total energy is just $F/(hL^2)=\eta\Delta f$, so that $\eta$ is either 0 or 1 depending on the sign of $\Delta f$. If $\Delta f>0$, the insulating phase is favored and $\eta=0$ while when the metal is favored $\Delta f<0$ and $\eta=1$; this homogeneous switching occurs at $\Delta f=0$, i.e. $T=T_c$.
\begin{figure}[t!]
  \includegraphics[width=\columnwidth]{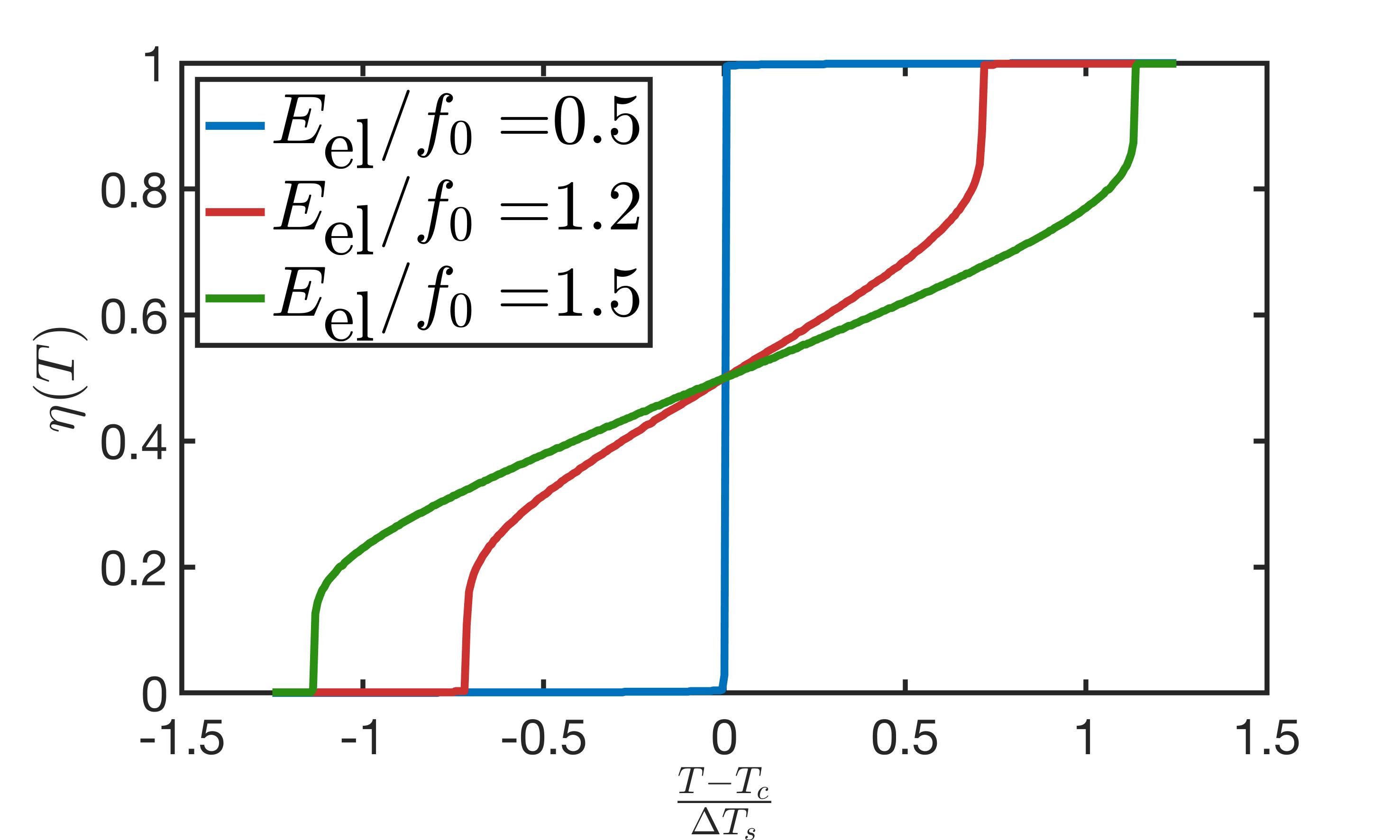}
  \caption{\scriptsize{Plot of $\eta(T)$ for three different values of $\Eel/f_0$ and for $\Edw/f_0=0.25$, $\nu=0.4$. For these parameters, the critical energy scale is $\Edw^c/\Eel\approx0.49$, so the blue curve $\Eel=0.5f_0=2\Edw$ is slightly above the threshold value and correctly exhibits a homogeneous phase switching at $T_c$.}}\label{F5}
\end{figure}

When $\Edw<\Edw^c$ we have to minimize numerically first over $q$ and then over the phase fraction to find $\eta(T)$, but we can make some general qualitative statements.

For the sake of simplicity we assume $\Delta f$ to be approximately linear in the temperature ($\Delta f\sim f_0\frac{T_c-T}{\Delta T_s}$) and $\Edw$ to be constant in $T$ (and crucially depending on the thickness through $\xi/h$). We first consider $\Delta f>0$; the opposite case is immediate given the symmetry around $\eta=1/2$ of the elastic energy.

The homogeneous $\eta=0$ insulating phase has zero energy, while a phase with $\eta=1/2$ gains a negative contribution $\sim-\Eel$ from elastic and domain wall energy and incurs a cost $\Delta f/2$ in electronic energy, so that if $\Delta f$ is large compared to $\Eel$ no stripe phase can exist. Since $\Delta f$ depends on $T$, small values of the ratio $\Eel/f_0$ suppress the formation of stripes to a narrow window of temperatures around $T_c$.

Thus we find that $\Eel$ has to be large enough compared to both $f_0$ and $\Edw$ in order to have formation of stripes. The ratio $\Eel/f_0$ determines the range of temperatures where a striped phase exists, while $\Eel/\Edw$ controls if stripes can appear at all: when this ratio is too low $\eta$ jumps from $0$ to $1$ at $T=T_c$, see Fig. \ref{F5}, and the system switches in bulk between homogeneous phases.

We observe that for intermediate values of $\Eel$, the phase fraction jumps discontinuously from 0 to a finite value and then evolves smoothly up to $1/2$ for $T=T_c$ (the behavior for $\eta>1/2$ is symmetric). This behavior is in agreement with experimental results in \CRObis \cite{CRO:237}. The dependence on $T$ between the two endpoint jumps is determined by the temperature dependence of $\Delta f$ and $\fb$, which is model specific.

\section{Metastable stripes}\label{MetaStab}

In this section we study more in detail the case of a stripe ordering extending only down to a certain depth $d$ smaller than the film thickness $h$. To keep things simple, we restrict to the isotropic case ($\sigma_{xx}=\sigma_{yy}$) of the previous sections, for a first order transition at the critical temperature $T=T_c$.

\subsection{Elastic free energy}\label{MetaStabEl}

We consider an order parameter constant in $z$ for $0<z<d$ and zero otherwise, i.e. $\psi_q\sim\Theta(d-z)$ (with $\Theta$ the step function). In such regime the particular solution is found using the known Green function of an infinite elastic medium, while the coefficients of the homogeneous solution are calculated in the same way as in Section \ref{Model} and Appendix \ref{App:A}, see Appendix \ref{App:ElEqdh} for more details.
\begin{figure}[t!]
  \centering
  \includegraphics[width=\columnwidth]{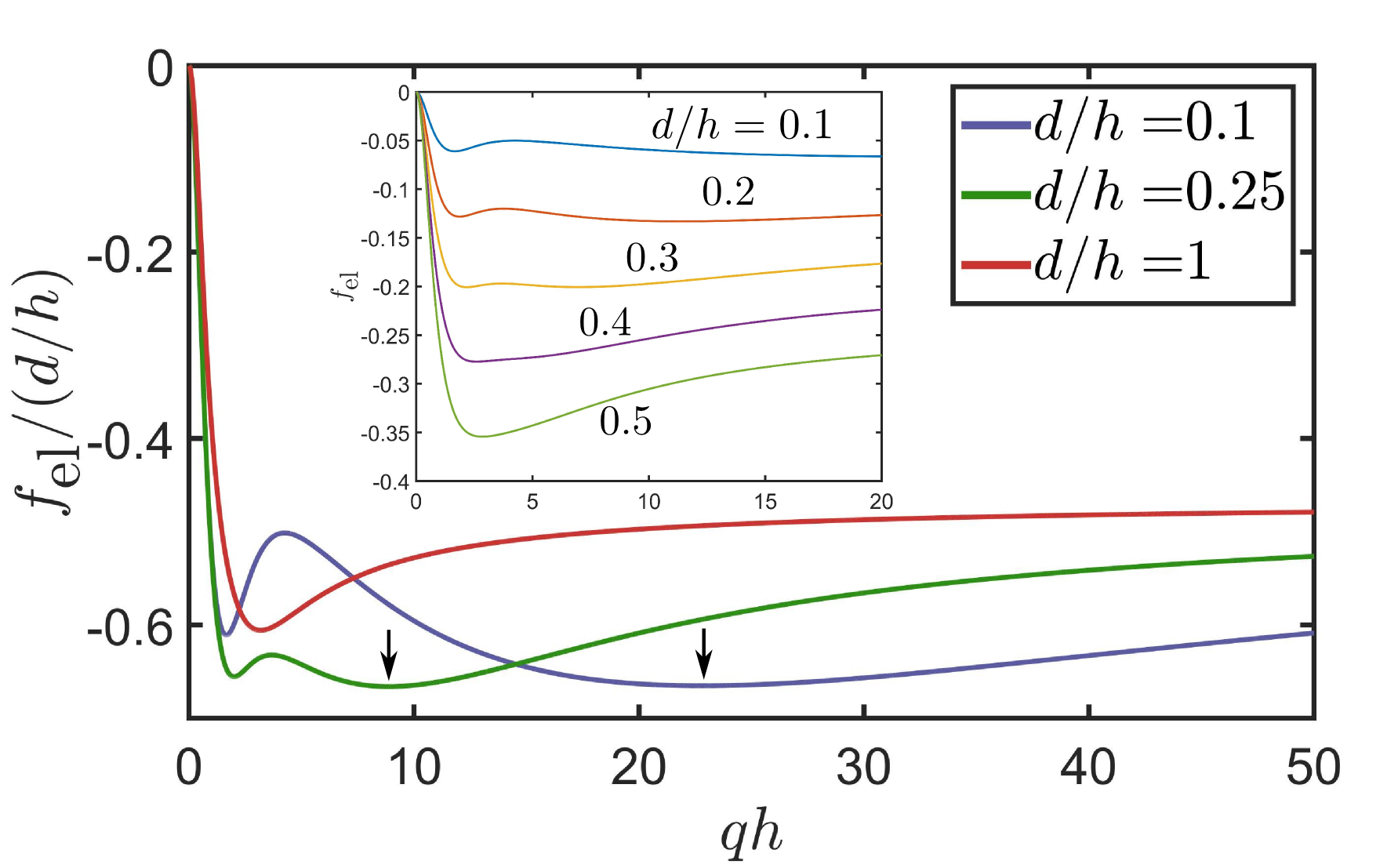}
  \caption{\scriptsize{Plot of the rescaled elastic energy $f_{\textrm{el}}/(d/h)$ for three different values of $d$ and for $\nu=0.4$; the inset shows the behavior of $f_{\textrm{el}}$ for $d/h=0.1,0.2,0.3,0.4,0.5$. The arrows indicate the position of the secondary minima at $qd\sim1$ for $d/h=0.1,\,0.25$.}}\label{F2}
\end{figure}

We find that the resulting free energy has the same qualitative features as the energy in Eqs. \eqref{FhasymInf}-\eqref{Fhasym0}, but is roughly rescaled by a factor $d/h$, i.e. $f_{\textrm{el}}$ is approximately proportional to $d$, see Fig. \ref{F2}; this is consistent with our arguments in Section \ref{ModelEqsSol}. Moreover, $\fel$ vanishes at $q=0$ and has a negative concavity independently of $d/h$:
\begin{gather}
\label{fnuhdq2}f_{\textrm{el}}(q\sim0)=-(1+\nu)\frac{(qh)^2}3\frac dh\left[3-\left(\frac dh\right)^2\right];\\
\label{fnuhdasy}f_{\textrm{el}}(q\rightarrow\infty)=-(1+\nu)\frac{1-2\nu}{1-\nu}\frac dh.
\end{gather}

An important qualitative difference compared to the $d=h$ case is the presence of a second minimum around $qd\sim1$, besides the minimum at $qh\sim1$, which is evident especially for $d\ll h$. The two minima have roughly comparable energy and are due to the presence of two lengthscales ($d$ and $h$); they merge into one minimum when the two lengths are comparable, i.e. for $d/h\gtrsim0.5$ (see the inset in Fig. \ref{F2}).

\subsection{Total free energy}

We now modify the calculations of Section \ref{PT1En} to account for $d$.

The order parameter in Eq. \eqref{OPpsi} becomes
\begin{gather}\label{OPpsid}
\psi(\mb x)=\psi_I+\Delta\psi\sum_n\phi(x-n\lambda)\Theta(d-z),
\end{gather}
where we set $\eta=1/2$ since we operate at $T=T_c$ and the variational parameters are $q$, $\ldw$ and $d$.

The contributions from the domain wall energy in Eqs. \eqref{Fgr} is rescaled by a factor $d/h$, so Eq. \eqref{Fgr2} becomes
\begin{equation}\label{Fgrd}
\frac{F_{\textrm{dw}}}{hL^2}=qd\frac{\ldw}hf_{\textrm b}+\frac{f_0\xi^2}{h\ldw}\Delta\psi^2qd.
\end{equation}

The domain wall thickness $\ldw$ minimizing Eq. \eqref{Fgrd} does not change, but we have to include the contribution from the domain walls in the $x$-$y$ plane, which appear when $d$ is different from $0$ or $h$ (more precisely when $d>\xi$ or $h-d<\xi$). For $\eta=1/2$ this additional term is equal to zero for $d=0$ or $d=h$ and $F_{\textrm{dw}}^x=L^2hE_{\textrm{dw}}/2$ otherwise. For practical purposes we employ a regularized form and write the total energy Eq. \eqref{FT} as
\begin{equation}\label{FTTcd}
\frac{F}{hL^2}=\Edw qd+E_{\textrm{el}}\sum_{q_m}f_{\textrm{el}}(q_m)|\psi_{q_m}|^2+\frac{\Edw}{2}w(d),
\end{equation}
where $w$ is a regularized function that vanishes at $0$ and $d=h$ and rapidly converges to $1$ for intermediate values.

\begin{figure*}[!ht]
  \includegraphics[width=\textwidth]{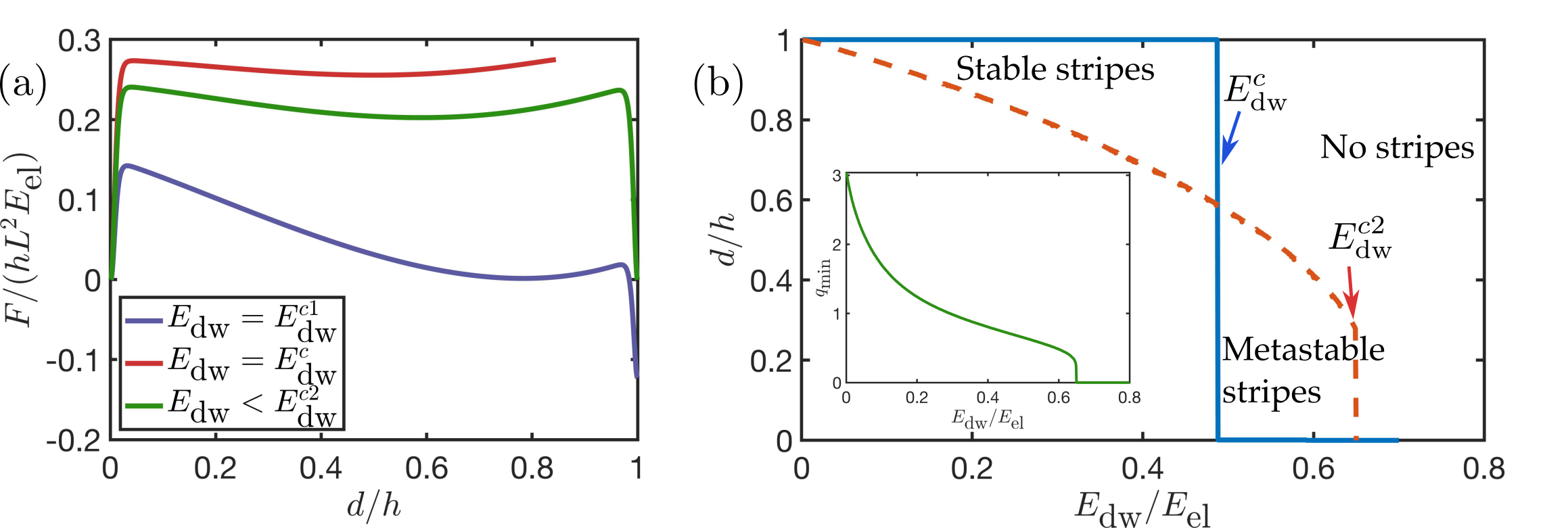}
  \caption{\scriptsize{(a) Plot of $F$ normalized to $\Eel hL^2$ as function of $d/h$ for $\nu=0.4$ and three values of the domain wall energy $\Edw=\Edw^{c1}\approx0.3\Eel$, $\Edw=\Edw^{c}\approx0.49\Eel$ and $\Edw=0.55\Eel$; the red curve corresponding to $\Edw=0.55\Eel$ does not extend up to $d=h$ because no minimum in $q$ of the free energy exists for $\Edw>\Edw^c$. (b) Plot of the values of $d/h$ for the globally stable striped phase (solid blue) and for the metastable striped phase (dashed red) as function of $\Edw/\Eel$; the values of $\Edw^c=0.49\Eel$ and $\Edw^{c2}=0.65\Eel$ are reported on the plot. The inset shows the optimal wavevector $\qmin$ as function of $\Edw/\Eel$.}}\label{F6}
\end{figure*}

The sum over $q_m$ is performed in the same way as for $d=h$, but its behavior at small $q$ becomes, using Eq. \eqref{fnuhdq2}:
\begin{equation}\label{FsumApd}
\Fel\sim\Eel\fel''(q=0)q/h\sim\Eel\frac12\left[3-\left(\frac dh\right)^2\right]qd.
\end{equation}
We observe that the threshold value of $\Edw$ required to form stripes depends on $d$, due to the $(d/h)^2$ correction in the small $q$ behavior of $\fel$, and is smaller compared to the $d=h$ case.

The extra energy cost due to the in-plane domain walls is substantial and such that the global minimum of the energy \eqref{FTTcd} is always located at $d=h$ or $d=0$. Nonetheless, a phase with stripes with $d<h$ may exist in a metastable form; in particular, from Eq. \eqref{FsumApd} we can write
\begin{equation}\label{Edwcd}
\Edw^c(d)=\Edw^c\frac12\left[3-\left(\frac dh\right)^2\right],
\end{equation}
and find that stripes with $d<h$ can exist even at $\Edw\geq\Edw^c$, while stripes with $d=h$ cannot. Such stripes are only metastable since their energy is still larger than the homogeneous $d=0$ phase.

We can minimize numerically the energy \eqref{FTTcd} with respect to $q$ and then with respect to $d$ to find the optimal thickness $d_{\textrm{min}}$ of the metastable stripes as function of $\Edw/\Eel$. Depending on the value of the energy at $d=d_{\textrm{min}}$ we find two more critical value of the domain wall energy.

For $\Edw<\Edw^{c1}$, $F(d=h)<F(d_{\textrm{min}})<F(d=0)$ and the stripe phase with $d=h$ is globally stable, while the homogeneous phase and the stripe phase with $d<h$ are locally stable, although the homogeneous phase has a larger energy. For $\Edw^{c1}<\Edw<\Edw^{c}$, $F(d=h)<F(d=0)<F(d_{\textrm{min}})$: the homogeneous phase has lower energy than the $d<h$ striped phase, but is still globally unfavored compared to the $d=h$ stripe phase. For $\Edw^{c}<\Edw<\Edw^{c2}$, the global minimum is the homogeneous phase and the $d=h$ stripe phase cannot exist: $F(d=0)<F(d_{\textrm{min}})$. Finally for $\Edw>\Edw^{c2}$ no stripes can exist in either metastable or stable forms. This behavior is depicted in Fig. \ref{F6}a; moreover Fig. \ref{F6}b shows the optimal thickness $d/h$ of the stripes: the solid line shows the global energy minimum which is either a $d=h$ striped phase below $\Edw^c$ or a homogeneous $d=0$ phase above $\Edw^c$; the dashed line shows the metastable $d<h$ striped phase.

The periodicity wavevector $\qmin$ (inset in Fig. \ref{F6}b) scales like $\sim1/h$ independently of $d$. This behavior is a consequence of the fact that the two minima of $\fel$ at $q\sim1/d$ and $q\sim1/h$ have roughly the same energy (see Fig. \ref{F2}), but the domain wall energy scales like $\sim qd$, so that its positive contribution is a factor $d/h$ smaller for the minimum at $q\sim1/h$, making it energetically favored.

Notice that the metastable striped phase is very difficult to access since it requires the control of $\Edw/\Eel$, which is mostly independent of easily controllable parameters such as the temperature. However, it may be possible to drive the system into such state by using an external pumping \cite{AbsNeg:20,ZhS:AJM1,ZhS:AJM2} that creates an energy landscape inhomogeneous in $z$, thus favoring a thinner striped phase like the driving current does in the experiments of Ref. \cite{Mengkun:CRO}. If parameters are finely tuned enough, the system may be driven into this metastable striped phase and remain trapped in it even after the removal of the pump \cite{ZhS:AJM1,ZhS:AJM2}.

\section{Orientation of stripes}\label{Orientation}

In this subsection we consider the question of the orientation of the stripes.

The elastic free energy can be expressed as $\fel\sim\sum f_{ij}\tilde\sigma_{ij}^2+f_{xx,zz}\tilde\sigma_{xx}\tilde\sigma_{zz}$, where $\tilde\sigma$ is the rotated stress tensor Eq. \eqref{Eq:sij1}, which depends on the angle $\theta$ between $\mb q$ and the $x$ axis.

In general the stripes are oriented in the direction that minimizes the combination of elastic and electronic energy. In low symmetry situations the orientation is determined by material properties and by the anisotropies of the coupling; the question becomes more tractable in higher-symmetry situations where interesting and more precise statements can be made. For simplicity, we also assume that the domain wall energy is small so that the periodicity of the stripes is essentially set by the behavior of $\fel$.

In the thin-film geometry of main interest in this paper, we may consider different levels of symmetry under rotations around an axis normal to the film. The simplest case is an electronic theory with a $C_4$ or higher rotational symmetry and an electronic order parameter that is invariant under rotations so that it couples to changes of the in-plane area of the unit cell. In this case the nonzero terms in the coupling tensor $\sigma_{ij}$ are $\sigma_{xx}=\sigma_{yy}=\sigma_0$ and $\sigma_{zz}$. In the  isotropic elasticity case considered here the elastic energy is rotationally invariant and the stripes may be oriented in any direction.
\begin{table}[t]
    \centering
    \begin{tabular}{|c|c|c|c|}
    \hline
    Symmetry & $\sigma_{ij}$ & Orientation & Periodicity\\
    \hline
        Isotropic & $\sigma\mathbb{1}_{xy}+\sigma_{zz}$ & No $\theta$ & $q=q(\nu)$\\
        \hline
        $C_4$ - JT & $\vec\sigma_{JT}\cdot\vec\tau_{xy}$ & $\pm45^{\circ}$ to princ. axes & $q\rightarrow\infty$\\
        \hline
        $C_4$ - $p$ & $\sigma_{xz}=\sigma_{yz}=\sigma_{p}$ & no stripes & $q=0$\\
        \hline
        $C_2$ & $\sigma\mathbb{1}_{xy}+\vec\sigma_{JT}\cdot\vec\tau_{xy}$ & $2$ minima $\rightarrow\theta_{\textrm{min}}$ & $q=q(\nu)$\\
        \hline
    \end{tabular}
    \caption{Table of the main symmetry classes of the coupling tensor $\sigma_{ij}$ and the resulting orientation and periodicity of the stripes.}
    \label{tab:Orientation}
\end{table}

One may next consider an electronic order parameter that preserves parity but changes sign under $\pi/2$ rotation,  spontaneously  breaking $C_4$ to $C_2$. For consistency with the notation in the rest of the paper we choose the principal axes of the electronic distortion such that the non-zero terms in the coupling tensor are $\sigma_{xy}=\sigma_{yx}=\sigma_{JT}$. In this situation Eq. \eqref{Eq:fel} becomes
\begin{equation}
\fel(q)=f_{xx}(q)\sin^2(2\theta)+f_{xy}(q)\cos^2(2\theta).
\label{Eq:fJT}
\end{equation}

A numerical inspection of $f_{xx}(q)$ shows that it is always larger than $f_{xy}(q)$, so that Eq. \eqref{Eq:fJT} is minimized at $\theta=0,\pi/2$ and at $q\rightarrow\infty$. In this case there are two stripe orientations, at a $\pm45^\circ$ angle to the principal axes defined by the electronic distortion, while the length scale is set by electron-scale physics. 
\begin{figure*}[t!]
  \centering
  \includegraphics[width=\textwidth]{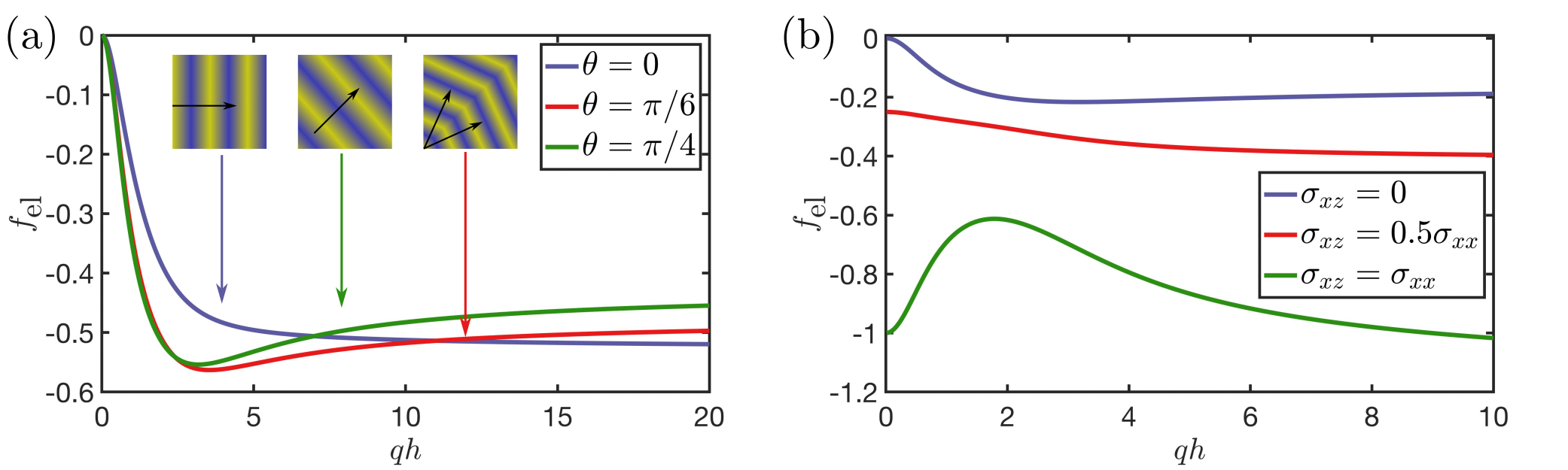}
  \caption{\scriptsize{(a) Plot of $f_{\textrm{el}}$ for three values of the angle $\theta$ between $\mb q$ and the $x$ axis, for $\nu=0.4$, $\sigma_{xz}=\sigma_{yz}=\sigma_{zz}=0$, $\sigma_{xx}=\sigma_{yy}$, $\sigma_{xy}=-0.6\sigma_{xx}$. (b) Plot of $f_{\textrm{el}}$, for three different values of $\sigma_{xz}$ (measured in units of $\sigma_{xx}$) and for $\nu=0.4$, $\sigma_{zz}=\sigma_{yz}=\sigma_{xy}=0$ and $\sigma_{xx}=\sigma_{yy}$.}}\label{Fthxz}
\end{figure*}

An electronic order parameter that breaks $C_4$ to $C_2$ and also breaks inversion symmetry can couple to $\epsilon_{xz,yz}$ via the coupling $\sigma_{xz}=\sigma_{yz}=\sigma_P$. Evaluation of Eq. \eqref{Eq:fel} for this case shows that the energy is $\fel\sim(f_{yz}-f_{xz})\sigma_P^2\sin2\theta$. An inspection of $f_{yz}(q)-f_{xz}(q)$ reveals that its minimum is at $q=0$, i.e. no stripes form because the bulk elastic distortion is compatible with the boundary conditions. 

Finally, if the electronic symmetry is $C_2$ instead of $C_4$ then within an appropriate choice of axes the allowed coupling terms are $\sigma_{xx}=\sigma_{yy}=\sigma$ and $\sigma_{xy}=\sigma_{yx}=\sigma_{JT}$. If $\sigma/\sigma_{JT}$ is sufficiently small, then there are two possible stripe directions $\theta_{\textrm{min}}$ and $\pi/2-\theta_{\textrm{min}}$; as the ratio increases the two directions collapse to one allowed direction at $\theta_{\textrm{min}}=45^\circ$, along the axis favored by the $C_2$ symmetry coupling. 

We get a quantitative verification by considering the regime $\nu\lesssim0.3$, where the minimum of $f_{xx}$ and thus $\fel$ is at $q\rightarrow\infty$, and the angle dependent terms are
\begin{gather}\label{Eq:Fth2}\fel\sim\frac{\sigma_{JT}}{1-\nu}\left(\frac12\sigma_{JT}\sin^22\theta+(1-2\nu)\sigma\sin2\theta\right);
\end{gather}
The minimum $\fel$ is at $\sin2\theta=-(1-2\nu)\sigma/\sigma_{JT}$, confirming the presence of two minima at $\theta_{\textrm{min}}=\text{arcsin}(-(1-2\nu)\sigma/\sigma_{JT})/2$ and $\pi/2-\theta_{\textrm{min}}$, which merge into one minimum at $\theta=\pi/4$ for $|\sigma/\sigma_{JT}|>1/(1-2\nu)$, see Fig. \ref{Fthxz}a.

We remark that, similarly to the considerations on checkerboard patterns in Section \ref{PT2}, the two optimal orientations at $\theta_{\textrm{min}}$ and $\pi/2-\theta_{\textrm{min}}$ do not result in a checkerboard pattern, since regions of such pattern would have unfavorable values of $\psi$. Instead, the stripes may arrange in ``macro-domains'' each with one of the two optimal orientations, see inset in Fig. \ref{Fthxz}a. The shape of the domains will depend on local defects and on the energy cost of the boundaries between these domains. This prediction explains the recent results of nanoimaging experiments on \CRObis \cite{CRO:237}, where similar macrodomains with different orientations have indeed been observed.

These considerations are summarized in Table~\ref{tab:Orientation}. They show how the presence of off-diagonal components in $\sigma_{ij}$ creates an optimal orientation for the stripes. The general case is usually very complicated, since a minimization over both $\theta$ and $q$ is usually required, but qualitatively similar.

We have also seen that the components of $\sigma_{ij}$ affect the optimal value of $q$ and the presence of stripes. Using the expression for $\fel$ derived in Appendix \ref{App:A}, we can actually write the quadratic coefficient in the small $q$ limit for the general case, and make precise predictions on the appearance of stripes.
\begin{eqnarray}\notag
c_{q^2}\tilde\sigma_{xx}^2 = &\frac23&\Big(\tilde\sigma_{xx}^2+\tilde\sigma_{xy}^2-\tilde\sigma_{yz}^2-\frac{4}{2(1-\nu)}\tilde\sigma_{xz}^2+\\
\label{Eq:felasyc2}&+&\frac{1-6\nu}{2(1-\nu)}\tilde\sigma_{zz}\tilde\sigma_{xx}-\frac{\nu(1-4\nu)}{2(1-\nu)^2}\tilde\sigma_{zz}^2\Big).
\end{eqnarray}

A necessary condition for stripes to appear is $c_{q^2}>0$, so all negative terms in Eq. \eqref{Eq:felasyc2} suppress the formation of stripes. It is the case of $\tilde\sigma_{xz}$, $\tilde\sigma_{yz}$ and $\tilde\sigma_{zz}$ (although in the latter case it  may depend on the specific value of $\nu$), while $\tilde\sigma_{xy}$ and $\tilde\sigma_{xx}$ favor the formation of stripes. Since the rotation mixes separately $\sigma_{xx}$ with $\sigma_{xy}$ and $\sigma_{xz}$ with $\sigma_{yz}$, we can state that any stress mismatch in the $x$-$y$ plane ($\sigma_{xx}$, $\sigma_{xy}$ or $\sigma_{yy}$) favors stripes, while shear distortions along $z$ ($\sigma_{xz}$ and $\sigma_{yz}$) suppress stripes, see Fig. \ref{Fthxz}b; finally distortions along $z$ ($\sigma_{zz}$) may either suppress or favor stripes depending on $\nu$ and the relative sign with $\sigma_{xx}$.

The suppressing influence of $\sigma_{xz}$ and $\sigma_{yz}$ agrees with what we have seen, since a homogeneous phase switching is compatible with the boundary conditions and thus is more favorable.

\section{Conclusions}\label{conclusions}

We studied a general model of an electronic system undergoing a metal-insulator phase transition and coupled to the elastic degrees of freedom of the lattice.

We wrote the elastic equations for a system with a finite thickness and solved them as function of the electronic order parameter for typical boundary conditions (stress-free surface at the top and clamped surface at the bottom).

Even considering a minimal model isotropic in the $x$-$y$ plane, we found that in most cases an inhomogeneous electronic phase (with a mixture of metallic and insulating stripes) lowers the elastic energy compared to an homogeneous phase; this is due to the presence of the boundary constraints which create a frustration in the elastic strain that results in homogeneous distortions being unfavorable. The energy gained scales quadratically with the electrons-lattice coupling parameter and is minimized for a wavevector approximately equal to the inverse of the system thickness.

We then minimized the total free energy in the case of a first order and a second order phase transition. In both cases we find that stripes emerge in the vicinity of the transition when the typical elastic energy gained by forming stripes is larger than the typical energy cost of a domain wall. We note that the ratio between elastic and domain wall energy is proportional to $\xi^2/h^2$ for second order transitions and to $\xi/h$ for first order transitions, i.e. a smaller mismatch tensor is required to form stripes in systems with a second order criticality. On the other hand, larger values of the electron-lattice coupling are typical of first order transitions in the electrons degrees of freedom, so that a model-specific analysis is needed to correctly assess this point and it is not possible to make general predictions. We considered a linear electron-lattice coupling, and a second order or weakly first order transition, meaning that our model does not directly apply to strongly non-nonlinear systems. The basic physics of competition between elastic drive, elastic boundary conditions, and domain wall energies should still apply, but the consequences could be quantitatively
different. Extension of our work to the strongly coupled or strong first order case is an
important open problem.

Furthermore we showed that stripes are the most stable when they extend across the entire depth of the system. This seems consistent with observations in \CRObis \cite{CRO:237}, but is somewhat in contrast to what reported in current-induced stripes in \CRO \cite{Mengkun:CRO}, where the stripe pattern is confined to a thin surface layer. The difference resides in the different mechanism driving the transition; in \CRObis the physics involves tuning through an equilibrium phase transition, in contrast to a nonequilibrium drive leading to a spatially dependent temperature variation in \CRO.

Nonetheless, even in homohgeneous situations, stripes extending only partway through the sample may still exist in a metastable form; they are difficult to access by homogeneously varying the temperature of the system, but may be investigated using inhomogeneous driving mechanisms, and the connection of these to the phenomena observed in the nonequilibrium experiments is an important open problem.

In the case of a first order transition we also investigated the evolution of the phase fraction of the stripe phase as function of the temperature in the spinodal region. We found that the temperature range of existence of the stripes depends on the ratio between the elastic energy and the difference in electronic energy between the metal and insulating phase, and showed that the temperature dependence of the phase fraction is in agreement with recent experimental results.

Finally we analyzed at a qualitative level the preferred orientation of stripes, finding that in general they have up to two possible directions, and showed that the appearance of stripes is suppressed by shear components of the stress in the $z$ direction. We also showed that when more than one orientation is possible, stripes arrange in unistripe macrodomains, rather than in a bistripe checkerboard pattern, providing an explanation for experimental observations in \CRObis. We did not investigate the problem of stripes orientation in detail, and reserve a more quantitative study for future applications of this model to specific materials.

\emph{Acknowledgements} -- We thank A. S. McLeod for useful comments and discussions. G. C. acknowledges the hospitality of Flatiron Institute.

\bibliographystyle{apsrev4-1}
\bibliography{Stripes_bib}

\onecolumngrid

\pagebreak[4]

\appendix

\section{Solution of the elastic equations in the general case}\label{App:A}

In this appendix we show in detail how to solve the elastic equations, finding the particular solution and the coefficients of the homogeneous solutions, and obtain the elastic free energy. The order parameter is constant in $z$ and $d=h$.

If the elastic tensor is isotropic, we have rotational invariance in the $x$-$y$ plane and we can rotate the axis to align $\mb q$ with the $x$ axis, i.e. choosing $q_y=0$. We call $\theta$ the angle between this new coordinate system and the principal axes. The components of $\sigma_{ij}$ will change because of this rotation and will be given by 
\begin{equation}\label{Eq:sij1}
\tilde\sigma_{ij}=
\begin{pmatrix}
\sigma_{xx}\cos^2\theta+\sigma_{yy}\sin^2\theta-\sigma_{xy}\sin2\theta & \sigma_{xy}\cos2\theta+(\sigma_{xx}-\sigma_{yy})\cos\theta\sin\theta & \sigma_{xz}\cos\theta-\sigma_{yz}\sin\theta\\
\sigma_{xy}\cos2\theta+(\sigma_{xx}-\sigma_{yy})\cos\theta\sin\theta & \sigma_{yy}\cos^2\theta+\sigma_{xx}\sin^2\theta+\sigma_{xy}\sin2\theta & \sigma_{yz}\cos\theta+\sigma_{xz}\sin\theta\\
\sigma_{xz}\cos\theta-\sigma_{yz}\sin\theta &  \sigma_{yz}\cos\theta+\sigma_{xz}\sin\theta & \sigma_{zz}
\end{pmatrix}
\end{equation}

\subsection{Solution of the elastic equations}\label{App:A1}

If we assume an order parameter constant in $z$, we find the particular solution at once by inverting Eq. \eqref{EqElIs} in Fourier space:
\begin{equation}\label{uP}
\vec u^P=\frac{2(1+\nu)}E\frac{\psi_{\mb q}}{iq}e^{iqx}\begin{pmatrix}
\frac{1-2\nu}{2(1-\nu)}\tilde\sigma_{xx}\\
\tilde\sigma_{xy}\\
\tilde\sigma_{xz}
\end{pmatrix}
\end{equation}

When $q_y=0$, only $u_2$ has components along $y$
\begin{gather}
\label{uHomY}u_0^{\pm}=e^{\mp qz}\left[\begin{pmatrix}1\\0\\0\end{pmatrix}\mp\frac{qz}{3-4\nu}\begin{pmatrix}1\\0\\\pm i\end{pmatrix}\right];\quad 
u_1^{\pm}=e^{\mp qz}\begin{pmatrix}1\\0\\\pm i\end{pmatrix};\quad
u_2^{\pm}=e^{\mp qz}\begin{pmatrix}0\\-1\\0\end{pmatrix};
\end{gather}
and we can effectively decouple the boundary conditions into a 4 by 4 system in the $x-z$ plane and a 2 by 2 system along the $y$ direction. We write $\vec A_{xz}\equiv(A_0^+,A_0^-,A_1^+,A_1^-)$ and $\vec A_y\equiv(A_2^+,A_2^-)$ and the system of equations originating from the boundary conditions reads
\begin{gather}
\label{AxzBxz}\vec A_{xz}\equiv \frac{2(1+\nu)}E\psi_{\mb q}M_{xz}^{-1}\cdot(\vec B_{xz}^P+\vec B_{xz}^{\psi});\qquad
\vec B_{xz}^{\psi}=\begin{pmatrix}
\tilde\sigma_{xz}\\
\tilde\sigma_{zz}\\
0\\
0
\end{pmatrix}; \qquad
\vec B_{xz}^P\equiv-\begin{pmatrix}
\tilde\sigma_{xz}\\
\frac{\nu}{1-\nu}\tilde\sigma_{xx}\\
\frac{(1-2\nu)\tilde\sigma_{xx}}{2iq(1-\nu)}\\
\frac{\tilde\sigma_{xz}}{iq}
\end{pmatrix};\\
\label{Mxz}M_{xz}\equiv\begin{pmatrix}
-2q & 2q & -q\frac{4(1-\nu)}{3-4\nu} & q\frac{4(1-\nu)}{3-4\nu}\\
-2iq & -2iq & -iq\frac{2(1-2\nu)}{3-4\nu} & -iq\frac{2(1-2\nu)}{3-4\nu}\\
e^{-qh} & e^{qh} & e^{-qh}\left(1-\frac{qh}{3-4\nu}\right) & e^{qh}\left(1+\frac{qh}{3-4\nu}\right)\\
ie^{-qh} & -ie^{qh} & -ie^{-qh}\frac{qh}{3-4\nu} & -ie^{qh}\frac{qh}{3-4\nu}
\end{pmatrix};\\
\label{AyMyBy}\vec A_y\equiv \frac{2(1+\nu)}E\psi_{\mb q}M_y^{-1}\cdot(\vec B_y^P+\vec B_y^{\psi});\qquad
\vec B_y^{\psi}=\begin{pmatrix}
\tilde\sigma_{yz}\\
0
\end{pmatrix}; \qquad
\vec B_y^P\equiv-\begin{pmatrix}
0\\
\frac{\tilde\sigma_{xy}}{iq}
\end{pmatrix}; \qquad
M_y\equiv\begin{pmatrix}
q & -q\\
-e^{-qh} & -e^{qh}
\end{pmatrix}.
\end{gather}

Solving the system we find
\begin{equation}\label{Ay}
\vec A_y=\frac{2(1+\nu)}E\psi_{\mb q}\frac{1}{2q\cosh(qh)}\begin{pmatrix}
e^{qh}\tilde\sigma_{yz}-i\tilde\sigma_{xy}\\
-e^{-qh}\tilde\sigma_{yz}-i\tilde\sigma_{xy}
\end{pmatrix};
\end{equation}
\begin{gather}\label{Axz}
\vec A_{xz}=\frac{2(1+\nu)}E\psi_{\mb q}\frac{a^{xx}\tilde\sigma_{xx}+a^{xz}\tilde\sigma_{xz}+a^{zz}\tilde\sigma_{zz}}{4q[(3-4\nu)\cosh(2qh)+5-12\nu+8\nu^2+2q^2h^2]};\\
\notag a^{xx}(s)\equiv\frac{-i}{1-\nu}\begin{pmatrix}
2\nu[(e^s+1)(3-4\nu)(1-\nu)-2s(1-\nu)+s^2]+(1-2\nu)[(4(1-\nu)(1-2\nu)-(3-4\nu)s)e^s-se^{-s}]\\
2\nu[(e^{-2s}+1)(3-4\nu)(1-\nu)+2s(1-\nu)+s^2]+(1-2\nu)[(4(1-\nu)(1-2\nu)+(3-4\nu)s)e^{-s}+se^s]\\
-(3-4\nu)[(1-2\nu)(e^{-s}+e^{s}(3-4\nu-2s)]+\nu(1-2s)+\nu(3-4\nu)e^{2s}]\\
-(3-4\nu)[(1-2\nu)(e^{s}+e^{-s}(3-4\nu+2s)]+\nu(1+2s)+\nu(3-4\nu)e^{-2s}]
\end{pmatrix};\\
\notag a^{xz}\equiv2e^{-qh}\begin{pmatrix}
-e^{2qh}(5-12\nu+8\nu^2+(3-4\nu)qh)-(3-4\nu-qh)\\
5-12\nu+8\nu^2-(3-4\nu)qh+e^{2qh}(3-4\nu+qh)\\
(3-4\nu)(1+(3-4\nu+2qh)e^{2qh})\\
-(3-4\nu)(3-4\nu+2qh+e^{2qh})\end{pmatrix};\\
\notag a^{xx}\equiv\frac{-i}{1-\nu}\begin{pmatrix}
2\nu[(e^{2qh}+1)(3-4\nu)(1-\nu)-2qh(1-\nu)+q^2h^2]+(1-2\nu)[(4(1-\nu)(1-2\nu)-(3-4\nu)qh)e^{qh}-qhe^{-qh}]\\
2\nu[(e^{-2qh}+1)(3-4\nu)(1-\nu)+2qh(1-\nu)+q^2h^2]+(1-2\nu)[(4(1-\nu)(1-2\nu)+(3-4\nu)qh)e^{-qh}+qhe^{qh}]\\
-(3-4\nu)[(1-2\nu)(e^{-qh}+e^{qh}(3-4\nu-2qh)]+\nu(1-2qh)+\nu(3-4\nu)e^{2qh}]\\
-(3-4\nu)[(1-2\nu)(e^{qh}+e^{-qh}(3-4\nu+2qh)]+\nu(1+2qh)+\nu(3-4\nu)e^{-2qh}]
\end{pmatrix};\\
\notag a^{xz}\equiv2e^{-qh}\begin{pmatrix}
-e^{2qh}(5-12\nu+8\nu^2+(3-4\nu)qh)-(3-4\nu-qh)\\
5-12\nu+8\nu^2-(3-4\nu)qh+e^{2qh}(3-4\nu+qh)\\
(3-4\nu)(1+(3-4\nu+2qh)e^{2qh})\\
-(3-4\nu)(3-4\nu+2qh+e^{2qh})\end{pmatrix};\\
\notag a^{zz}\equiv i\begin{pmatrix}
2((1-\nu)(3-4\nu)(1+e^{2qh})-2qh(1-\nu)+q^2h^2\\
2((1-\nu)(3-4\nu)(1+e^{-2qh})+e^{-2qh}(2qh(1-\nu)+q^2h^2)\\
-(3-4\nu)(1-2qh+(3-4\nu)e^{2qh})\\
-(3-4\nu)((3-4\nu)e^{-2qh}+1+2qh)\end{pmatrix}
\end{gather}

\subsection{Elastic energy}\label{App:A2}

The elastic energy for component $\psi_q$ can be written as
\begin{gather}
\notag\Fel=-\frac12\text{Re}\int d\mb x\sigma_{ij}\epsilon_{ij}\psi_qe^{-iqx}=-\frac12\psi_q\text{Re}\int d\mb x[iq(\tilde\sigma_{xx}u_x+\tilde\sigma_{xy}u_y+\tilde\sigma_{xz}u_z)+\partial_z(\tilde\sigma_{xz}u_x+\tilde\sigma_{yz}u_y+\tilde\sigma_{zz}u_z)];\\
\label{Eq:Fel_u}\Fel=-\frac{L^2}2\psi_q\text{Re}[iq\int dz(\tilde\sigma_{xx}u_x+\tilde\sigma_{xy}u_y+\tilde\sigma_{xz}u_z)-(\tilde\sigma_{xz}u_x+\tilde\sigma_{yz}u_y+\tilde\sigma_{zz}u_z)_{|z=0}],
\end{gather}
where we integrated by parts the $z$ derivative term and used that $\vec u(z=h)=0$. We observe that the particular solution is purely imaginary and proportional to $1/iq$, so it does not contribute to the free energy from the displacement at $z=0$, and gives a constant contribution from the first term. We can thus split the free energy into a contribution from the particular solution, which is independent of $q$, and a $q$-dependent term arising from the homogeneous solution:
\begin{gather}\notag
\Fel=-\frac{L^2}2\Big[\frac{2(1+\nu)}{E}\psi_q^2(\frac{1-2\nu}{2(1-\nu)}\tilde\sigma_{xx}^2+\tilde\sigma_{xy}^2+\tilde\sigma_{xz}^2)+\\
\notag+\psi_q\text{Re}[iq(\tilde\sigma_{xx}\vec C_{x}\cdot\vec A_{xz}+\tilde\sigma_{xy}\vec C_{y}\cdot\vec A_{y}+\tilde\sigma_{xz}\vec C_{z}\cdot\vec A_{xz})-(\tilde\sigma_{xz}\vec C_{0x}\cdot\vec A_{xz}+\tilde\sigma_{yz}\vec C_{0y}\cdot\vec A_{y}+\tilde\sigma_{zz}\vec C_{0z}\cdot\vec A_{xz})]\Big],
\end{gather}

where $\vec C_{0i}$ arise from evaluating the homogeneous solutions at $z=0$ and are
\begin{equation}\label{C0i}
\vec C_{0x}=(1,1,1,1); \quad \vec C_{0y}=(-1,-1);\quad \vec C_{0z}=(0,0,i,-i),
\end{equation}

while $\vec C_i$ come from integrating the solutions over $z$:
\begin{gather}
\label{Cx}\vec C_{x}=\frac1q(1-e^{-qh}-\frac{1-(1+qh)e^{-qh}}{3-4\nu},e^{qh}-1+\frac{1+(qh-1)e^{qh}}{3-4\nu},1-e^{-qh},e^{qh}-1);\\
\label{CyCz}\vec C_{y}=-\frac1q(1-e^{-qh},e^{qh}-1);\qquad
\vec C_{z}=\frac i q(-\frac{1-(1+qh)e^{-qh}}{3-4\nu},-\frac{1+(qh-1)e^{qh}}{3-4\nu},1-e^{-qh},1-e^{qh}).
\end{gather}

We only retain the terms that have a non zero imaginary part, group them up and write
\begin{gather}\notag
\frac{\Fel}{hL^2}=-\frac12\frac{2(1+\nu)}{E}\psi_q^2\Big[\frac{1-2\nu}{2(1-\nu)}\tilde\sigma_{xx}^2+\tilde\sigma_{xy}^2+\tilde\sigma_{xz}^2+\frac{\sinh(qh)}{qh\cosh(qh)}(\tilde\sigma_{yz}^2-\tilde\sigma_{xy}^2)+\\
\notag+\frac1{4qh[(3-4\nu)\cosh(2qh)+5-12\nu+8\nu^2+2q^2h^2]}\text{Re}[\tilde\sigma_{xx}^2iq\vec C_{x}\cdot\vec a^{xx}+\\
\notag+\sigma_{xz}^2(iq\vec C_{z}-\vec C_{0x})\cdot\vec a^{xz}+\tilde\sigma_{xx}\tilde\sigma_{zz}(iq\vec C_{x}\cdot\vec a^{zz}-\vec C_{0z}\cdot\vec a^{xx})+\tilde\sigma_{zz}^2(-\vec C_{0z})\cdot\vec a^{zz}]\Big].
\end{gather}

Explicitly
\begin{gather}
\label{Eq:FEL}
\frac{\Fel}{hL^2}=\frac{\tilde\sigma_{xx}^2}{E}\sum_q\fel(q)|\psi_q|^2;\\
\label{Eq:fel}
\fel(q)\tilde\sigma_{xx}^2=(1+\nu)[f_{xx}\tilde\sigma_{xx}^2+f_{xy}\tilde\sigma_{xy}^2+f_{xz}\tilde\sigma_{xz}^2+f_{yz}\tilde\sigma_{yz}^2+f_{zz}\tilde\sigma_{zz}^2+f_{xx,zz}\tilde\sigma_{zz}\tilde\sigma_{xx}];\\
\label{Eq:fxx}
f_{xx}=\frac{(1-4\nu+\nu^2+4\nu^3)\text{sh}(2qh)-4\nu(1-2\nu)[(1-2\nu)\text{sh}(qh)-qh\text{ch}(qh)]+2qh(1-4\nu+5\nu^2)}{qh(1-\nu)[(3-4\nu)\text{ch}(2qh)+5-12\nu+8\nu^2+2q^2h^2]}-\frac{1-2\nu}{2(1-\nu)};\\
\label{Eq:fxyfyz}
f_{xy}=\frac{\tanh(qh)}{qh}-1;\qquad f_{yz}=-\frac{\tanh(qh)}{qh};\\
\label{Eq:fxz}
f_{xz}=4(1-\nu)\frac{\sinh(2qh)-2qh}{qh[(3-4\nu)\cosh(2qh)+5-12\nu+8\nu^2+2q^2h^2]}-1;\\
\label{Eq:fxxzz}
f_{xx,zz}=2\frac{\nu(3-4\nu)\sinh(2qh)-2\nu qh+2(1-2\nu)^2\sinh(qh)-2(1-2\nu)qh\cosh(qh)}{qh[(3-4\nu)\cosh(2qh)+5-12\nu+8\nu^2+2q^2h^2]};\\
\label{Eq:fzz}
f_{zz}=-(1-\nu)\frac{(3-4\nu)\sinh(2qh)-2qh}{qh[(3-4\nu)\cosh(2qh)+5-12\nu+8\nu^2+2q^2h^2]}.
\end{gather}

Combining Eq. \eqref{Eq:sij1} with Eqs. \eqref{Eq:FEL}-\eqref{Eq:fzz}, one can minimize the elastic free energy with respect to $\theta$ and find the optimal orientation of the stripes.

\section{Elastic energy for a square wave order parameter}\label{App:sumfel}

From Eq. \eqref{FT} we write the elastic energy for a square wave-like order parameter as
\begin{equation}\label{AppFel}
\Fel=hL^2E_{\textrm{el}}\Delta\psi^2\sum_mf_{\textrm{el}}(q_m)\frac{2\sin^2(\pi\eta m)}{\pi^2 m^2}.
\end{equation}

Notice that for $\eta=1/2$, the Fourier component reduces to $2\Delta\psi^2/\pi^2m^2$ for odd $m$ and zero otherwise.

For small $q$, we can again approximate $\fel$ with $\sim q_m^2$, so that we calculate something like
\begin{equation}\label{AppFelq0}
\Fel\sim\sum_{m\lesssim1/qh}\frac{m^2q^2\sin^2(\pi\eta m)}{m^2}\sim\frac12q,
\end{equation}
where we used the fact that for $\eta$ not too close to $0$ or $1$ we are just averaging $\sin^2$ over many periods. Therefore the small $q$ behavior of the elastic energy is independent of $\eta$ for most values of the phase fraction. When $\eta$ goes to $0$ or $1$ the elastic energy vanishes, and we observe that the magnitude of $\Fel$ decreases (symmetrically) as $\eta$ moves away from $1/2$. 

\begin{figure}
    \centering
    \includegraphics[width=0.9\textwidth]{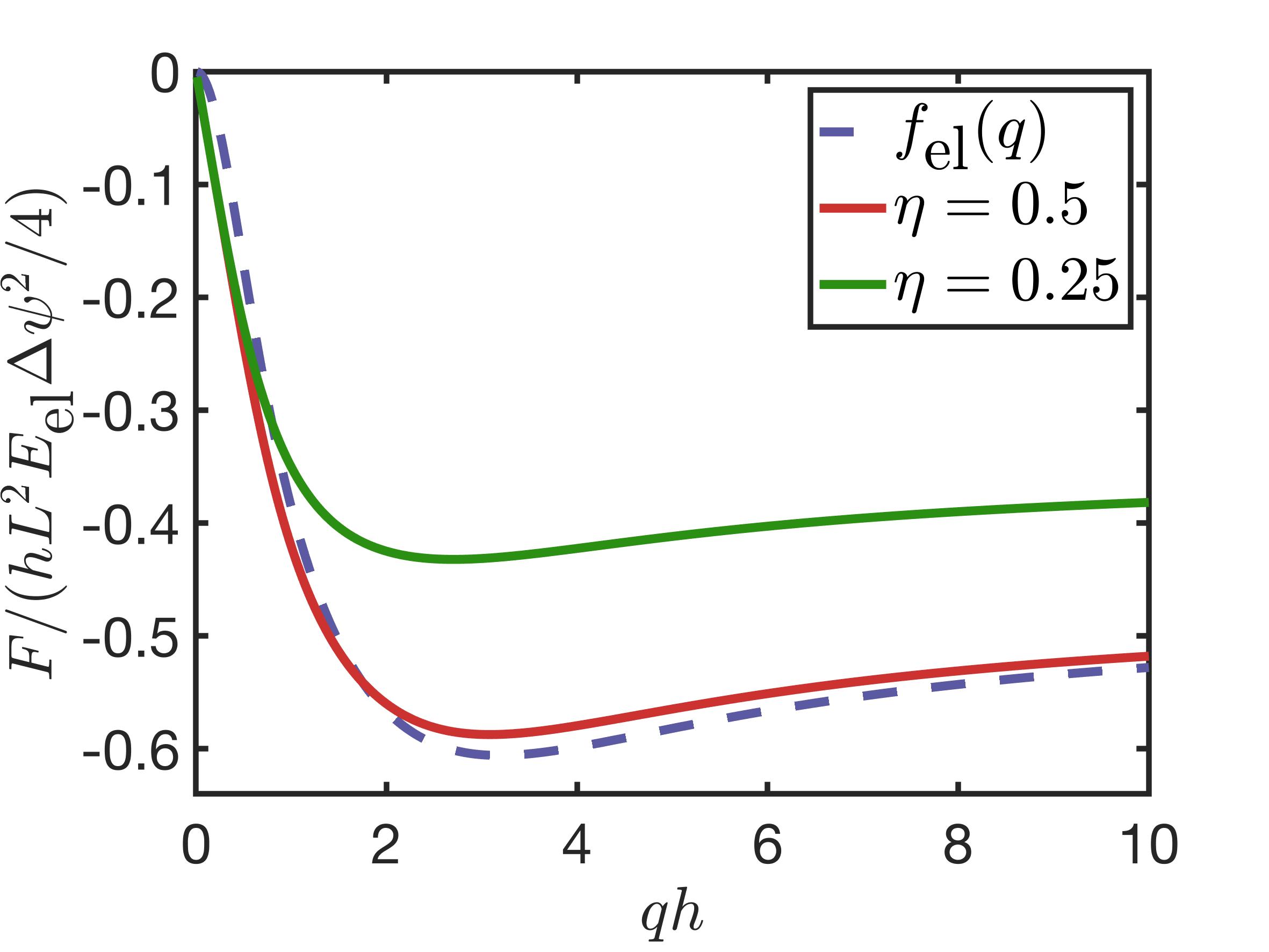}
    \caption{\scriptsize{Plot of the normalized elastic energy for a harmonic order parameter (dashed blue) and for a square wave like order parameter (red for $\eta=0.5$ and green for $\eta=0.25$) as function of the wavevector $q$, for $d=h$ and $\nu=0.4$.}}
    \label{FAppD1}
\end{figure}

These behaviors are shown in Fig. \ref{FAppD1}, where we also plot $\sum_m\fel|\psi_m|^2$ for two values of $\eta$, compared to the elastic $\fel$ for a purely harmonic order parameter $\psi\sim e^{iqx}$.

We observe the different behavior at small $q$: quadratic for a harmonic order parameter and linear for a square wave order parameter; we also notice that the elastic energy for a square wave does not depend on $\eta$ for small $q$, as predicted.

\section{Solution of the elastic equation for $d<h$}\label{App:ElEqdh}

If the striped phase occupies a thin ordered layer in the region $0<z<d$, we can write the electronic order parameter as $\psi=\psi_0e^{iqx}\Theta(d-z)-\psi_0\Theta(z-d)$, where for simplicity $\pm\psi_0$ are the values of the electronic order parameter that correspond to the two competing phases.

The particular solution can be found by employing the Green function in the $x$-$z$ plane
\begin{equation}\label{G2D}
G(q,q_z)=\frac1{\mu}\left(\frac{\delta_{\alpha\beta}}{q^2+q_z^2}-\frac{1}{2(1-\nu)}\frac1{(q^2+q_z^2)^2}
\begin{pmatrix}
  q^2 & qq_z \\
  qq_z & q_z^2
\end{pmatrix}\right),
\end{equation}
and noticing that the Fourier transform of the $x$ derivative of the order parameter is $iq\psi(q_z)=iq\psi_0ie^{-iq_zd}/(q_z+ i\delta)$, with $\delta\rightarrow0$ to ensure the regularity of the solution at $z\rightarrow-\infty$. The particular solution in real space for $z$ is then found as (in the regime $\sigma_{xx}=\sigma_{yy}$ and the other $\sigma_{ij}=0$)
\begin{equation}\label{uP2D}
\begin{pmatrix}
u_x^P\\
u_z^P
\end{pmatrix}=\int\frac{dq_z}{2\pi}e^{iq_zz}(-G_{\alpha\beta})
\begin{pmatrix}
\sigma_{xx}iq\psi(q_z)\\
0
\end{pmatrix},
\end{equation}
where we are assuming for simplicity that $\sigma_{xz}=\sigma_{zz}=0$. Lifting this assumption would lead to the appearance of additional terms in the integral for $u_P$, including $\delta(z-d)$ terms originating from the $z$ derivative of $\psi$. Equation \eqref{uP2D} can be computed using the residues method and finding
\begin{gather}\label{uxP2D}
u_x^P=\frac{-iq\sigma_{xx}\psi_0e^{iqx}}{\mu}\frac1{2q^2}\left[\frac{1-2\nu}{1-\nu}\theta(d-z)+\text{sgn}(z-d)e^{-q|z-d|}\left(1-\frac{2+q|z-d|}{4(1-\nu)}\right)\right];\\
\label{uzP2D}u_z^P=\frac{iq\sigma_{xx}\psi_0e^{iqx}}{\mu}\frac{i}{2q^2}e^{-q|z-d|}\frac{1+q|z-d|}{4(1-\nu)}.
\end{gather}

We now need to follow the same procedure as in Eqs. \eqref{AxzBxz}-\eqref{AyMyBy} to impose the boundary conditions and obtain the homogeneous solution; notice that now $\vec B_{\psi}=0$, while $\vec B_P$ is determined by Eqs. \eqref{uxP2D} and \eqref{uzP2D}.

The expression for the displacement $\vec u$ can then be substituted into the formula for the free energy. The calculations are rather cumbersome, but can be carried out analytically using a calculus software such as Mathematica; for book-keeping reasons we do not report them here.

\end{document}